\documentclass[12pt]{article}
\usepackage{amsmath,amssymb,amsfonts,amscd,amsxtra}
\usepackage{latexsym}

\input{epsf.sty}
\usepackage{epsfig}
\usepackage{graphicx,color}
\usepackage{graphics}
\usepackage{subfigure}
\usepackage{appendix}
\usepackage[top=1in, bottom=1in, left=1in, right=1in]{geometry}

\renewcommand{\theequation}{\thesection.\arabic{equation}}

\newtheorem{rem}{Remark}

\renewcommand{\Im}{{\mbox{Im}}}
\renewcommand{\Re}{{\mbox{Re}}}

\renewcommand\appendix{\par
  \setcounter{section}{0}
  \setcounter{subsection}{0}
  \setcounter{figure}{0}
  \setcounter{table}{0}
  \renewcommand\thesection{Appendix \Alph{section}}
  \renewcommand\theequation{\Alph{section}.\arabic{equation}}
  \renewcommand\thefigure{\Alph{section}.\arabic{figure}}
  \renewcommand\thetable{\Alph{section}.\arabic{table}}
  \renewcommand\thethm{\Alph{section}.\arabic{thm}}
}
\DeclareRobustCommand{\rchi}{{\mathpalette\irchi\relax}}
\newcommand{\irchi}[2]{\raisebox{\depth}{$#1\chi$}} 

\numberwithin{equation}{section}

\date{}

\title{A super-resolution imaging approach via subwavelength hole resonances}
 \author{
Junshan Lin
\thanks{Department of Mathematics and Statistics, Auburn University, Auburn, AL 36849 (jzl0097@
auburn.edu). Junshan Lin was partially supported by the NSF grant DMS-1719851.}
and Hai Zhang
\thanks{Department of Mathematics, 
HKUST,  Clear Water Bay, Kowloon, Hong Kong (haizhang@ust.hk). Hai Zhang was supported by HK RGC grant GRF 16304517 and GRF 16306318.}
}

\begin{document}

\maketitle

\begin{abstract}
This work presents a new super-resolution imaging approach by using subwavelength hole resonances.
We employ a subwavelength structure in which
an array of tiny holes are etched in a metallic slab with the neighboring distance $\ell$ that is smaller than half of the wavelength.
By tuning the incident wave at resonant frequencies, the subwavelength structure 
generates strong illumination patterns that are able to probe both low and high spatial frequency components of the imaging sample sitting above the structure.
The image of the sample is obtained by 
performing stable numerical reconstruction from the far-field measurement of the diffracted wave.
It is demonstrated that a resolution of $\ell/2$ can be obtained for reconstructed images,
thus one can achieve super-resolution by arranging multiple holes within one wavelength.

The proposed approach may find applications in wave-based imaging such as 
electromagnetic and ultrasound imaging.
It attains two advantages that are important for practical realization.
It avoids the difficulty to control the distance the between the probe and the sample surface with high precision.
In addition, the numerical reconstructed images are very stable against noise by only using the low frequency band of the far-field data
in the numerical reconstruction.
\end{abstract}

\setcounter{equation}{0}
\setlength{\arraycolsep}{0.25em}

\section{Introduction}
\label{sec:introduction}
Due to the wave nature of the optical light,
the resolution of conventional microscopies is typically constrained by 
the Rayleigh criterion (or \textit{Abbe's diffraction limit})
 \cite{Abbe, NB, Rayleigh}.
Enormous efforts have been devoted to achieve images beyond the diffraction limit in the last several decades.
One main class of microscopies achieve super-resolution by \textit{near-field} technologies, where one
probes the samples or collects the diffracted optical field in the vicinity of the object (typically within one wavelength) \cite{CB, CSS, Dunn, RWCSF}.
The principle of high resolution is provided by taking into account of evanescent waves, which decay exponentially at the object's surface.
The main limitation of such microscopies, however, lies in need to control the distance of the probe over the sample surface with extremely high precision and very shallow penetration depth of optical field. 
Therefore, there has been intensive efforts to explore super-resolution techniques without relying on near-field illumination/measurement.
Among those there are two main successful approaches.
One is the fluorescence type which relies on emission from fluorescent molecules and photon switching to localize a single molecular \cite{Betzig, HGM, RBZ}. The other is to use patterned/structured illumination to provide high spatial frequency wave pattern so as to shrink the support of point spread function \cite{Gustafasson1, Gustafasson2, Hell}.
However, so far microscopies based on structured illumination can only improve the Abbe's diffraction limit by a factor of two \cite{Gustafasson1}.

Mathematically,  the inverse scattering theory can be applied to understand how the structure of a scattering object is encoded in measured
wave fields in the near-field scanning optical microscopy and  other related imaging problems \cite{BLi2, BLi3, BL2, BL3, CMS, CS1, CS2, Chen}.
Numerical approaches have also been explored extensively to achieve super-resolution imaging of point sources and other related problems
when only far-field data is collected (cf. \cite{Candes, Demanet, Donoho, LF, Liu-Zhang} and references therein).

\subsection{The roadmap for super-resolution imaging}
Motivated by our recent work on the quantitative analysis of the  resonances for various subwavelength hole structures \cite{lin_shipman_zhang} - \cite{lin_zhang20_1}, 
in this paper we propose a new super-resolution imaging modality with illumination patterns generated by a collection of coupled subwavelength holes.
The original work of subwavelength hole array and the induced extraordinary optical transmission (EOT) dates back to about two decades ago \cite{ELGTW},
and this type of nanostructures has found important applications in biological and chemical sensing, and other novel optical devices; see, for instance, \cite{Blanchard-Meunier, Huang, Oh_Altug}.

The new illumination patterns are the resonant modes of the hole structures which oscillate on a subwavelength scale. They
can probe both the low and high spatial frequency components of the sample, and very importantly,  transfer such information to the lower frequency band of the wave field after its interaction with the sample and then propagates to the far-field detector plane. As such the high spatial frequency components of the sample
can be recovered through numerical inversion of the data measured in the far-field and one can achieve the super-resolution image for the sample. 
We would like to point out that subwavelenghth resonant modes using Helmholtz resonators can also give rise to
the super-focusing/super-resolution \cite{Lemoult-2011, Habib-Zhang}. 

\begin{figure}[!htbp]
\begin{center}
  \centering
  \includegraphics[height=3cm]{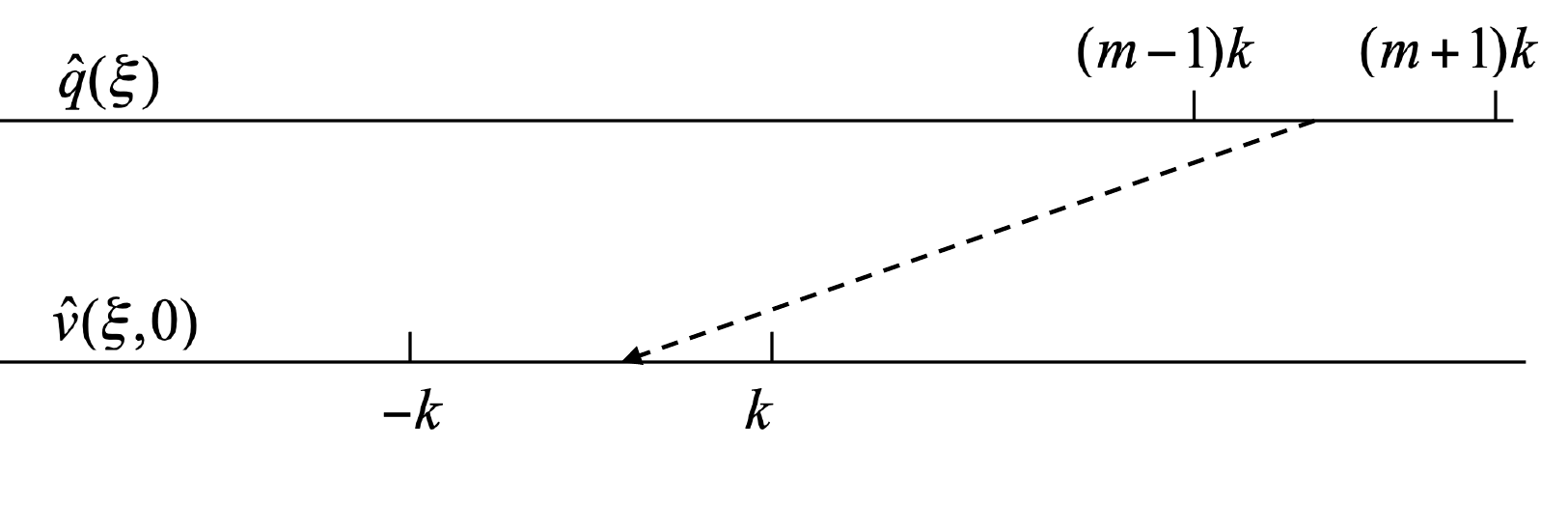}   
  \vspace*{-20pt}
\caption{Shift of $\hat{q}_{m}(\xi-mk)$ to  $\hat{v}_{m}(\xi,0)$ for $\xi\in\Omega_k$ when interacting with the illumination pattern $u_m$. }
  \label{fig:freq_shift}
\end{center}
\end{figure}

Let us describe the main idea for the above roadmap in the ideal case over the $x_1x_2$-plane. 
Suppose one would like to image a one-dimensional sample characterized by its transmission function $q(x_1)$ and the sample sits on the $x_1$-axis.
Let $\lambda$ be the free-space wavelength of the 
optical wave, and $k=2\pi/\lambda$ be the corresponding wavenumber. 
Consider illuminating the sample from below with a pattern given by $u_m=e^{i mkx_1 -\zeta x_2}$ for an integer $m$,
in which $(mk)^2 - \zeta^2 = k^2$ such that $u_m$ satisfies the frequency domain wave equation with the wavenumber $k$.
Then under the assumption that the interaction between the illumination and the sample is negligible,
the field through the sample becomes $v_m(x_1,0):=u_m(x_1,0) \cdot q(x_1)$, which then propagates to the detector plane $x_2=d \gg \lambda$.
Note that only the low frequency components of $v_m(x_1, 0)$ can reach the far-field detector plane
and be collected. They correspond to all the modes with spatial frequencies $\xi \in \Omega_k :=(-k, k)$.
Hence in the ideal scenario when the full-aperture data is available,
$\hat{v}_{m}(\xi,0)$ for $\xi \in \Omega_k$ can be recovered from the far-field measurement in the Fourier domain. Now the formula
\begin{equation}\label{eq:freq_shift}
\hat{v}_{m}(\xi,0) = \hat{u}_{m}(\xi,0) * \hat{q}(\xi) = \hat{q}(\xi-mk) \quad \mbox{for} \; \xi \in \Omega_k
\end{equation}
implies that the Fourier components $\hat{q}(\xi)$ for $\xi \in mk + \Omega_k$ is recovered from the far-field data (see Figure \ref{fig:freq_shift}).
Therefore, by varying $m$ from $-M$ to $M$, one recovers all the Fourier components of the sample transmission $q$ in the frequency band $[-(M+1)k, (M+1)k]$.

In this paper we employ the subwavelength holes to generate wave patterns that mimic the desired oscillations of
the ideal illuminations $u_m=e^{i mkx_1- \zeta x_2}$. These can be realized by arranging the subwavelength holes in an array and 
tuning the incidence at corresponding resonant frequencies as discussed in what follows.

\subsection{The proposed imaging setup and an overview of the imaging approach}
\begin{figure}[!htbp]
\begin{center}
 \centering
\includegraphics[height=7cm]{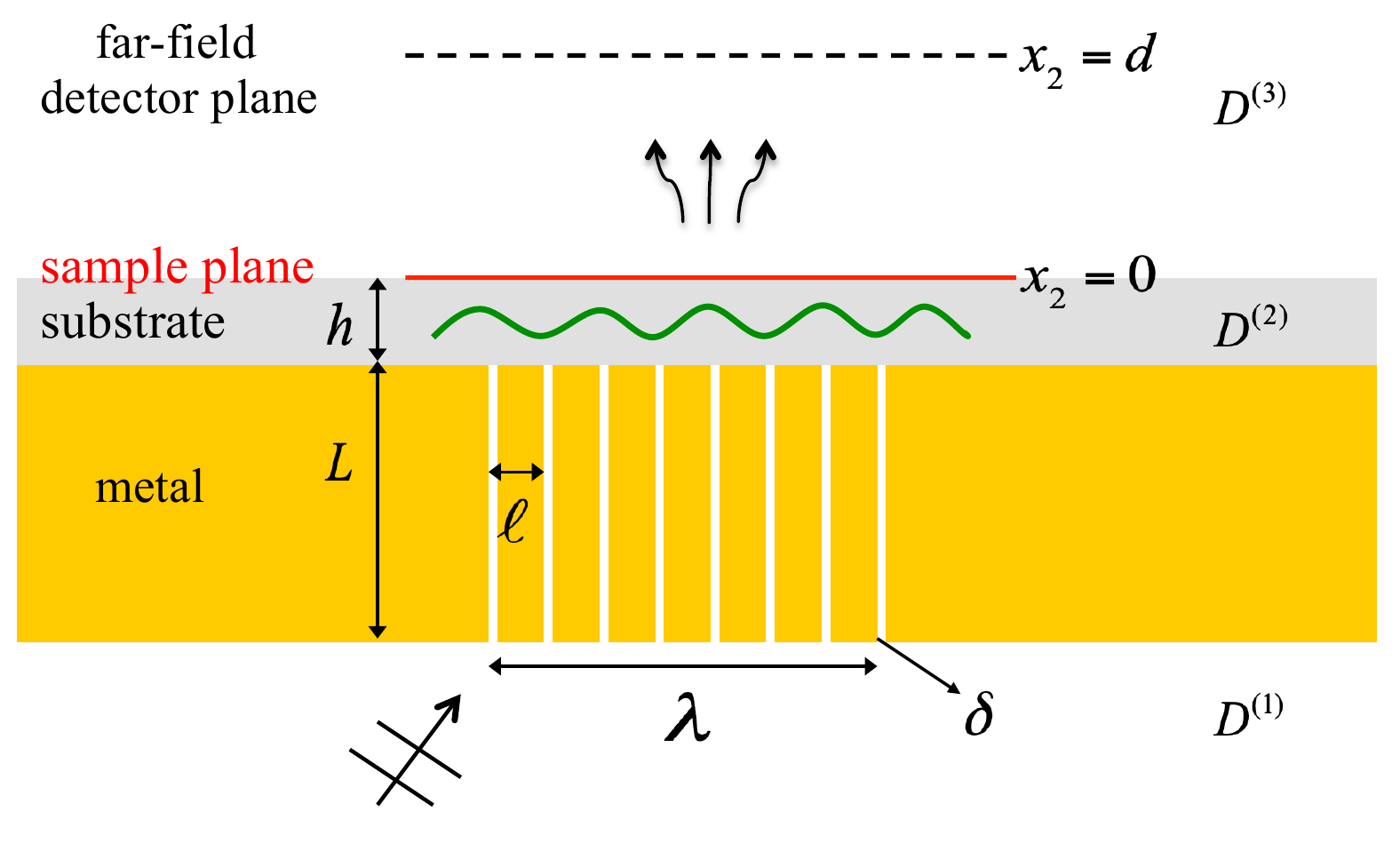} 
\caption{Schematic plot of the imaging setup on the $x_1x_2$-plane. }
  \label{fig:slits_imaging}
  \vspace*{-10pt}
\end{center}
\end{figure}

We focus on the imaging problem in the two-dimensional configuration, in which
the subwavelength structure consists of an array of identical slit holes $S_1, S_2, \cdots, S_J$ 
patterned in a metallic slab. The slits are invariant along the $x_3$ direction and Figure \ref{fig:slits_imaging} shows a schematic plot of the imaging setup on the $x_1x_2$-plane. 
The slits are arranged in a manner such that $\ell < \lambda/2$ and $J\ell = O (\lambda)$,
where $\ell$ denotes the distance between the adjacent slits, or more precisely, the distance between the left walls of two adjacent slit holes (see Figure \ref{fig:slits_imaging}).
In addition,  each slit hole has a width of $\delta$ and there holds $\delta \ll \ell$. The imaging sample is deposited over a substrate (e.g., glass) sitting on top of the metallic slab.
When an incident wave impinges from below the slab, it transmits through the slit holes and generates a wave pattern
that interacts with the sample. The corresponding diffracted field is then collected on the far-field detector plane.

For clarity of exposition, throughout the paper we assume that the slab and the substrate attains a thickness of $L=1$ and $h < \lambda$ respectively.
We adopt the coordinate system on the $x_1x_2$-plane such that the origin is located on the sample plane.
The support of the imaging sample on the sample plane is denoted by $I_0=(0, a)$. 
Let $a_1$ and $a_J$ be the $x_1$-coordinate of the left wall of $S_1$ and the right wall of $S_J$ respectively.
It is assumed that $I_0 \subset I_\mathrm{slit} :=[a_1,a_J]$ such that the support of the imaging sample is covered by the region that the slit apertures span.
For simplicity we assume that $a_1=0$.

\begin{rem}
The substrate is introduced here for the purpose of practical realization of the imaging setup. It also controls the near field interaction between the slit holes and the sample. 
Following the studies in \cite{lin_oh_zhang20}, one needs to have $h >\delta$ so that this interaction induces little shift for the resonant frequencies of the slit holes. On the other hand, smaller $h$ allows more interactions which may results higher resolution in the reconstructed image. 
We emphasize that  the inclusion of the substrate does not induce essential difference for the mathematical modeling of the imaging problem and its numerical reconstruction. 
The proposed imaging setup does not require controlling the distance between the probe and the sample surface with high precision as
in near-field microscopies.
\end{rem}

\medskip

The subwavelength structure attains a series of complex-values resonances lying below the real axis.
At the resonant frequencies (the real part of complex-valued resonances),
the transmission through the slit holes will exhibit peak values and the transmitted wave field is strong. In the periodic case, 
 almost total transmission can be achieved at the resonant frequencies \cite{lin_zhang18_1,lin_zhang18_2}.  
By tuning the frequency of the incidence field at several resonant frequencies only, the transmitted wave patterns will sweep from low to high frequencies, which allows for probing both the low and high spatial frequency components of the sample.
The resonances and wave patterns at resonant frequencies will be investigated in details in Section \ref{sec:resonance_structure}.

To obtain the sample image in a realistic configuration where only limited-aperture data is available,
one can not perform the reconstruction in the Fourier domain directly. 
Instead, we formulate and solve the underlying inverse problem in the spatial domain and present numerical algorithms to perform the reconstruction.
This will be elaborated in Section \ref{sec:inf-thin-sample}, where two approaches: one is based on
the gradient descent method and the other on the total variation regularization with the split Bregman iteration are presented.
One important feature of our numerical reconstruction for the underlying imaging problem is that
they are stable against noise, since only the lower frequency band $\Omega_k$ of the measured data is used, while the noise is typically highly oscillatory. 
This will be illustrated in Section \ref{sec:inf-thin-sample} and \ref{sec:thin-sample}  when the numerical algorithms and the numerical examples are presented.

In view of the relation \eqref{eq:freq_shift}, the resolution of the image depends on the oscillation patterns of the illumination, which is further determined by the distance $\ell$ between two adjacent slit holes in the proposed imaging setup. 
From various numerical examples given in Section \ref{sec:inf-thin-sample} and \ref{sec:thin-sample}, we observe that the image resolution is about $\ell/2$. 
Therefore, a high-resolution image can be obtained by arranging multiple holes within one wavelength.

\section{Resonant scattering by a collection of subwavelength holes}\label{sec:resonance_structure}
We formulate the mathematical model in the context of transverse magnetic (TM) electromagnetic wave scattering. 
The same model holds for acoustic waves, where one replaces the optical refractive index by the acoustic refractive index.
The domain below the metallic slab, the substrate domain and the domain above the substrate is denoted by $D^{(1)}$, $D^{(2)}$ and $D^{(3)}$, respectively.
For each slit hole $S_j$, let $\Gamma_{j}^{(1)}$ and $\Gamma_{j}^{(2)}$ denote the lower and upper slit apertures, respectively.
Let $D_\delta:=\cup_{j=1}^J S_j$ be the slit region, and
 $\Gamma^{(1)}$ and $\Gamma^{(2)}$ be the union of the lower and upper slit apertures $\Gamma_{j}^{(1)}$
and $\Gamma_{j}^{(2)}$ respectively. 
Then the relative permittivity $\varepsilon$ is given by
\begin{equation*}
\varepsilon(x)= \left\{
\begin{array}{lll}
\medskip
\varepsilon_0=1  & \mbox{for} \; x\in D^{(1)} \cup D^{(3)} \cup D_\delta, \\
\medskip
\varepsilon_h>1  & \mbox{for} \;  x\in D^{(2)},
\end{array}
\right.
\end{equation*}
and the refractive index value is $n(x)=\sqrt{\varepsilon(x)}$.

For the transverse magnetic polarization with the magnetic field $H=(0, 0, u)$, the Maxwell's equations
reduce to the scalar Helmholtz equation in two dimensions.
Let  $u_\mathrm{inc} = e^{i k( x_1\sin\theta \,+\, x_2\cos \theta )}$ be the incident plane wave that impinges from below the slab.
Denote the exterior region of the metal by $D$.
Then the total field $u$ satisfies
\begin{equation} \label{eq-scattering1}
\left\{
\begin{array}{llll}
\vspace*{0.1cm}
\nabla \cdot \left(\dfrac{1}{\varepsilon(x)} u \right) + k^2 u = 0   \quad\quad  \mbox{in} \; D, \\
\vspace*{0.1cm}
\dfrac{\partial u}{\partial\nu} = 0  \quad \mbox{on} \; \partial  D, \\
\vspace*{0.1cm}
[u] = 0,  \left[\dfrac{1}{\varepsilon} \dfrac{\partial u}{\partial \nu}\right] = 0    \quad\quad  \mbox{on} \; \partial D^{(3)} \cup\Gamma^{(2)}.
\end{array}
\right.
\end{equation}
In the above, $[\cdot]$ denotes the jump of the quantity when the limit is taken along the positive and negative unit normal direction $\nu$.
In addition, the diffracted field $u_\mathrm{diff}:=u-u^\mathrm{inc}$ satisfies outgoing radiation conditions at infinity.

It can be shown that the scattering problem \eqref{eq-scattering1} attains a unique solution for all complex wavenumber $k$ with $\Im k \geq 0$.
Moreover, the resolvent for the corresponding differential operator will attain a countable number of poles when continued meromorphically to the whole complex plane.
These poles are called the resonances (or scattering resonances) of the scattering problem, and the associated
resonant states (quasi-normal modes)  decay in time but grow exponentially away from the slab.

To obtain the resonances, we consider the problem \eqref{eq-scattering1} when the incident wave $u_\mathrm{inc}=0$ 
and set up an integral equation formulation \cite{Habib}. 
Let $ g^{(1)}(x,y)$ be the Green's function in the domain $D^{(1)}$ 
with the Neumann boundary condition along metallic slab boundary. Applying the Green's formula in $D^{(1)}$ gives 
$$u(x) = -\sum_{j=1}^J  \int_{\Gamma_{j}^{(1)}} g^{(1)}(x,y) \, \partial_2 u(y_-)  ds_y, \quad x\in D^{(1)}.$$
Here and henceforth, $\partial_2 u(y_{\pm})$ denotes the limit of the given function $\frac{\partial u}{\partial x_2}(x)$ when $x$ approaches the aperture from the above and below respectively.
Similarly, using the layered Green's function $g^{(2)}(x,y)$ in the domain $D^{(2)}\cup D^{(3)}$
with the Neumann boundary condition along metallic slab boundary, one obtains
$$u(x) = \sum_{j=1}^J \int_{\Gamma_{j}^{(2)}} g^{(2)}(x,y) \,  \partial_2 u(y_+)  ds_y, \quad x\in D^{(2)}.  $$
The derivation of the Green's function in layered medium is given in the appendix.
Let 
$$g_j(x,y):=g_0(x_1  - (j-1)\ell, x_2; y_1 - (j-1)\ell, y_2)$$ 
be the Green's function inside the slit $S_j$ with Neumann boundary condition along the boundary of $S_j$,
in which
$$ g_0(k;x,y)= \sum_{m,n=0}^\infty c_{mn}\phi_{mn}(x)\phi_{mn}(y),$$
with $c_{mn}\!=\![k^2\!-\!(m\pi/\delta)^2\!-\!(n\pi)^2]^{-1}$,
$ \phi_{mn}(x)\!=\!\sqrt{\frac{\beta_{mn}}{\delta}}\cos\left(\frac{m\pi}{\delta} x_1\right) \cos(n\pi x_2)$, and
\begin{equation*}
\beta_{mn} = \left\{
\begin{array}{llll}
1  & m=n=0, \\
2  & m=0, n\ge 1 \quad \mbox{or} \quad n=0, m\ge 1, \\
4  & m\ge 1, n \ge 1.
\end{array}
\right.
\end{equation*}
Then the solution inside the slit $S_j$ can be expressed as
$$ u(x) = \int_{\Gamma_{j}^{(1)}} g_j(x,y) \, \partial_2 u (y_+)  ds_y -  \int_{\Gamma_{j}^{(2)}}  g_j(x,y) \partial_2 u (y_-) ds_y $$ 
 for $x\in S_j$. 

By taking the limit of the above integral to the slit apertures and imposing the continuity condition of the electromagnetic field over the slit apertures,
we obtain the following system of boundary integral equations for $j=1, 2, \cdots, J$:
\begin{equation} \label{eq-scattering2}
\left\{
\begin{array}{llll}
& \hspace*{-10pt}   \displaystyle{\sum_{j=1}^J \int_{\Gamma_{j}^{(1)}}}  g^{(1)}(x,y) \varphi_j^{(1)}(y) ds_y  
+ \int_{\Gamma_{j}^{(1)}} g_j(x,y) \varphi_j^{(1)}(y)  ds_y \\
& + \displaystyle \int_{\Gamma_{j}^{(2)}} g_j(x,y) \varphi_j^{(2)}(y)  ds_y=0 \quad \mbox{on} \,\, \Gamma_{j}^{(1)},  \quad\quad\quad\quad\; (3) \\
& \hspace*{-10pt}   \varepsilon_h \displaystyle{\sum_{j=1}^J\int_{\Gamma_{j}^{(2)}}}  g^{(2)}(x,y) \varphi_j^{(2)}(y) ds_y  
+   \int_{\Gamma_{j}^{(1)}} g_j(x,y) \varphi_j^{(1)}(y)  ds_y \\
& +   \displaystyle \int_{\Gamma_{j}^{(2)}} g_j(x,y) \varphi_j^{(2)}(y)  ds_y
=0\quad \mbox{on} \,\, \Gamma_{j}^{(2)}, \nonumber
\end{array}
\right.
\end{equation}
where $\varphi_j^{(1)} := -\left. \partial_2 u (y_+) \right|_{\Gamma_{j}^{(1)}} = -\left. \partial_2 u (y_-) \right|_{\Gamma_{j}^{(1)}}$ 
and
$\varphi_j^{(2)} := \left. \partial_2 u (y_-) \right|_{\Gamma_{j}^{(2)}} = \frac{1}{\varepsilon_h}\left. \partial_2 u (y_+) \right|_{\Gamma_{j}^{(2)}}$.

The resonances are the characteristic values $k$ of the above integral equations for which nontrivial solutions $\{ \varphi_j^{(1)}, \varphi_j^{(2)} \}_{j=1}^J$ exist.
When $J=1$, the analytical expression for the resonances can be obtained through the asymptotic analysis of the integral equation and 
the Gohberg-Sigal theory. It can be shown that (cf. \cite{lin_oh_zhang20}) that the resonances obtain the following asymptotic expansion for 
integer $m$ satisfying $m\delta \ll 1$:
\begin{equation*}
k^{(m)} =  m \pi + m (\varepsilon_h+1) \cdot \delta \ln\delta + 2m\pi \cdot c_{m,H} \cdot \delta  + O(\delta^2\ln^2\delta),
\end{equation*}
where the complex-valued constant $c_{m,H}=O(1)$ is independent of $\delta$. Namely, the real part of the resonances is close to $m\pi$ and
their imaginary parts are of order $O(\delta)$.

When $J>1$, the coupling of the subwavelength holes will generate a group of resonances 
$\{ k_{m1}, \cdots,  k_{mJ} \}$ for each $m$ satisfying $m\delta \ll 1$.  The existence of resonances in this scenario
 can still be proved rigorously by applying the asymptotic expansion for \eqref{eq-scattering2} and the Gohberg-Sigal theory.
 This would boil down to solving for the roots of nonlinear functions $\gamma_1(k)$, $\cdots$, $\gamma_J(k)$, where each would attain a simple root
 near $m\pi$ for each given integer $m$. We refer the reader to \cite{lin_shipman_zhang} for such a rigorous analysis when $J=2$. 
Here we obtain the resonances by solving $\eqref{eq-scattering2}$ numerically. The computational approach we adopt
 follows the lines in \cite{lin_zhang19_1}, where a high-order numerical discretization of the integral operators as well as their fast implementation is achieved by 
a combination of the Nystrom scheme for singular kernels, a contour integration approach for the layered Green's function $g^{(2)}(x,y)$, and 
a Kummer's transformation acceleration strategy for evaluation of the series in the Green's function $g_j(x,y)$.
For a subwavelength structure with $\ell=0.2$ and $\delta=0.02$, the resonances with $J=3$ and $J=6$ are collected in 
Table \ref{tab:J=3} and \ref{tab:J=6} respectively for $m=1, 2, 3$. Here we set the substrate thickness as $h=0.1$, and its relative permittivity as $\varepsilon_h=2$.

\begin{table}[htbp]
\begin{center}
\caption{Resonances with $J=3$, $\ell=0.2$ and $\delta=0.02$.}
\medskip  
\begin{tabular}{ccccccc}
\hline
     & $k_{1,j}$   &  $k_{2,j}$    & $k_{3,j}$  \\ 
\hline
\hline    
$j=1$ & $2.8620 - 0.1791i$  & $5.7941 - 0.1412i$ & $8.7233 - 0.0949i$   \\
$j=2$ &  $2.8977 - 0.0193i$ & $5.8625 - 0.0131i$ & $8.8395 - 0.3822i$ \\
$j=3$ &  $2.9454 - 0.0013i$  & $5.9447 - 0.2737i$ & $9.0325 - 0.2863i$ \\
\hline
\end{tabular}
\label{tab:J=3}
\vspace*{-20pt}
\end{center}
\end{table}

\begin{table}[htbp]
\begin{center}
\caption{Resonances with $J=6$, $\ell=0.2$ and $\delta=0.02$.}
\medskip  
\begin{tabular}{ccccccc}
\hline
     & $k_{1,j}$   &  $k_{2,j}$    & $k_{3,j}$  \\ 
\hline
\hline    
$j=1$ & $2.8080 - 0.1186i$  & $5.6934 - 0.2542i$ & $8.6576 - 0.1087i;$   \\
$j=2$ & $2.8826 - 0.0146i$  & $5.7850 - 0.0472i$ & $8.6678 - 0.4521i$ \\
$j=3$ & $2.9213 - 0.2423i$  & $5.8559 - 0.0087i$ & $8.7540 - 0.0135i$ \\
$j=4$ & $2.9248 - 0.0032i$  & $5.8862 - 0.0014i$ & $9.0356 - 0.3623i$   \\
$j=5$ & $2.9439 - 0.0008i$  & $5.9533 - 0.3190i$ & $9.0368 - 0.2838i$ \\
$j=6$ & $2.9529 - 0.0002i$  & $6.0067 - 0.2431i$ & $9.0381 - 0.2680i$ \\
\hline
\end{tabular}
\label{tab:J=6}
\vspace*{-10pt}
\end{center}
\end{table}

\begin{figure}[!htbp]
\begin{center}
\hspace*{-0.35cm}
\includegraphics[height=3.5cm]{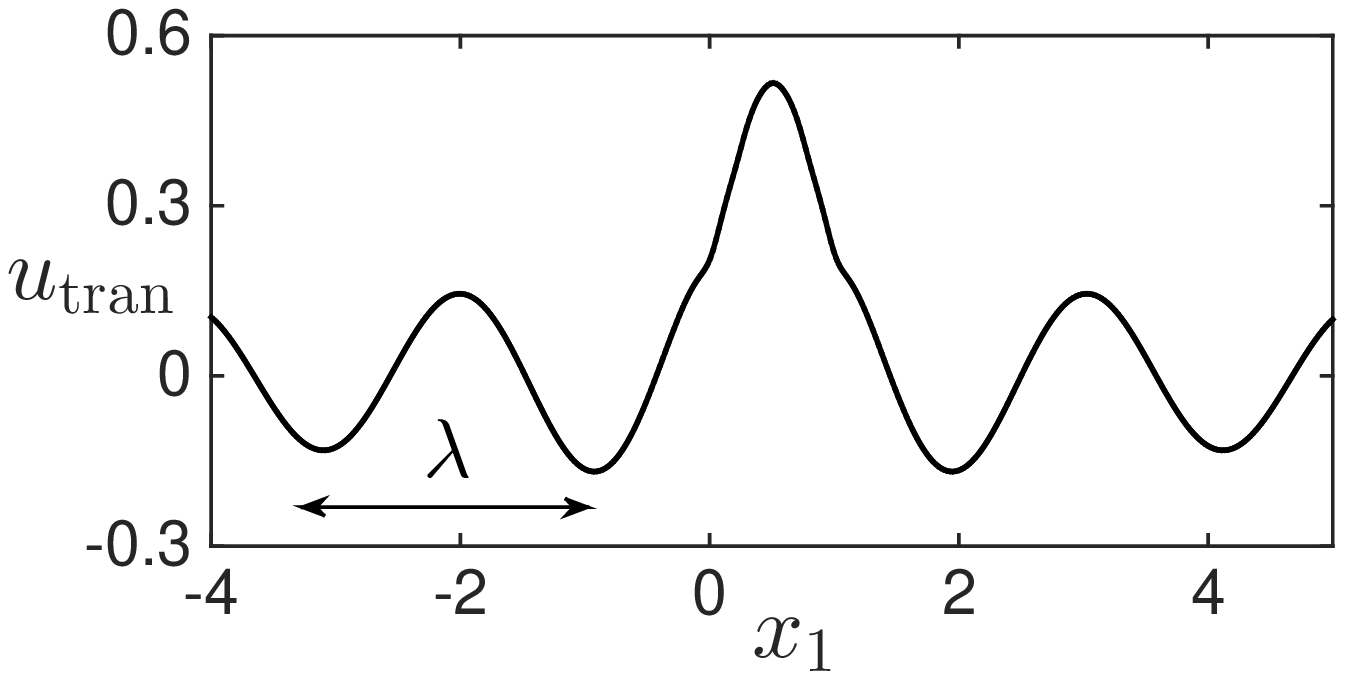} \hspace*{-0.5cm}
\includegraphics[height=3.5cm]{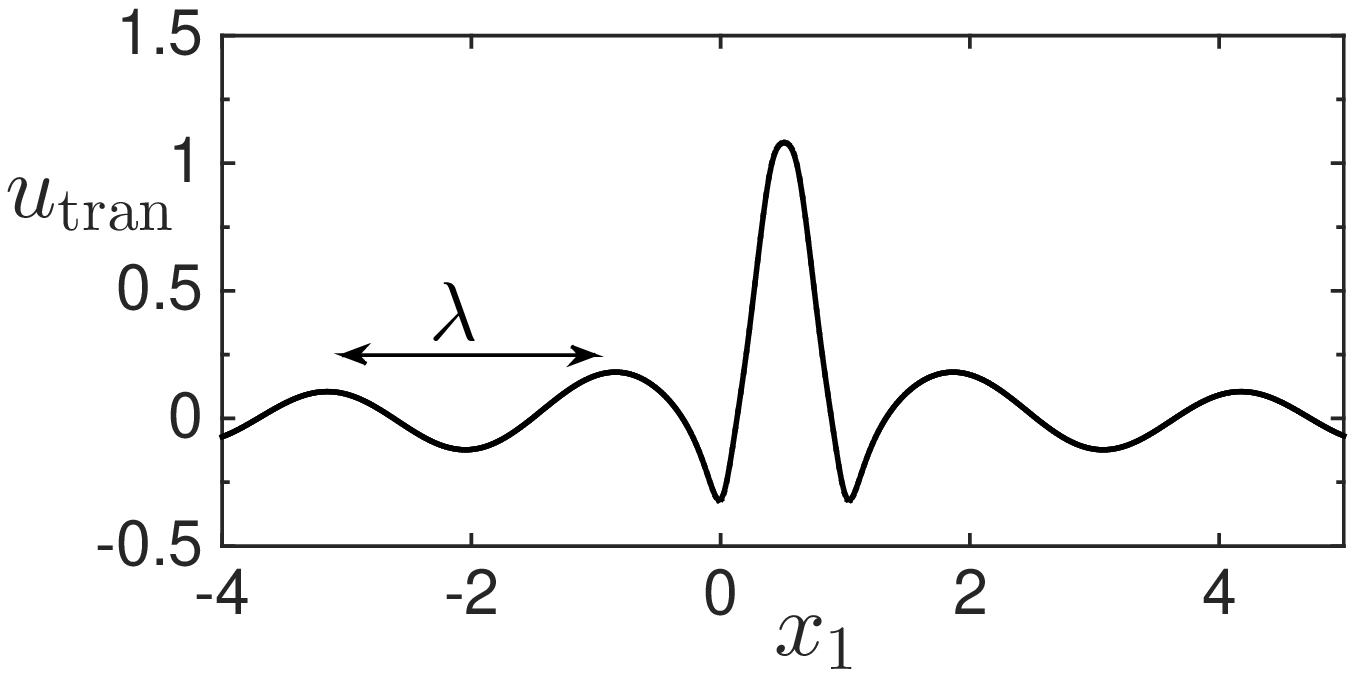} \hspace*{-0.5cm} \\
\hspace*{-0.35cm}
\includegraphics[height=3.5cm]{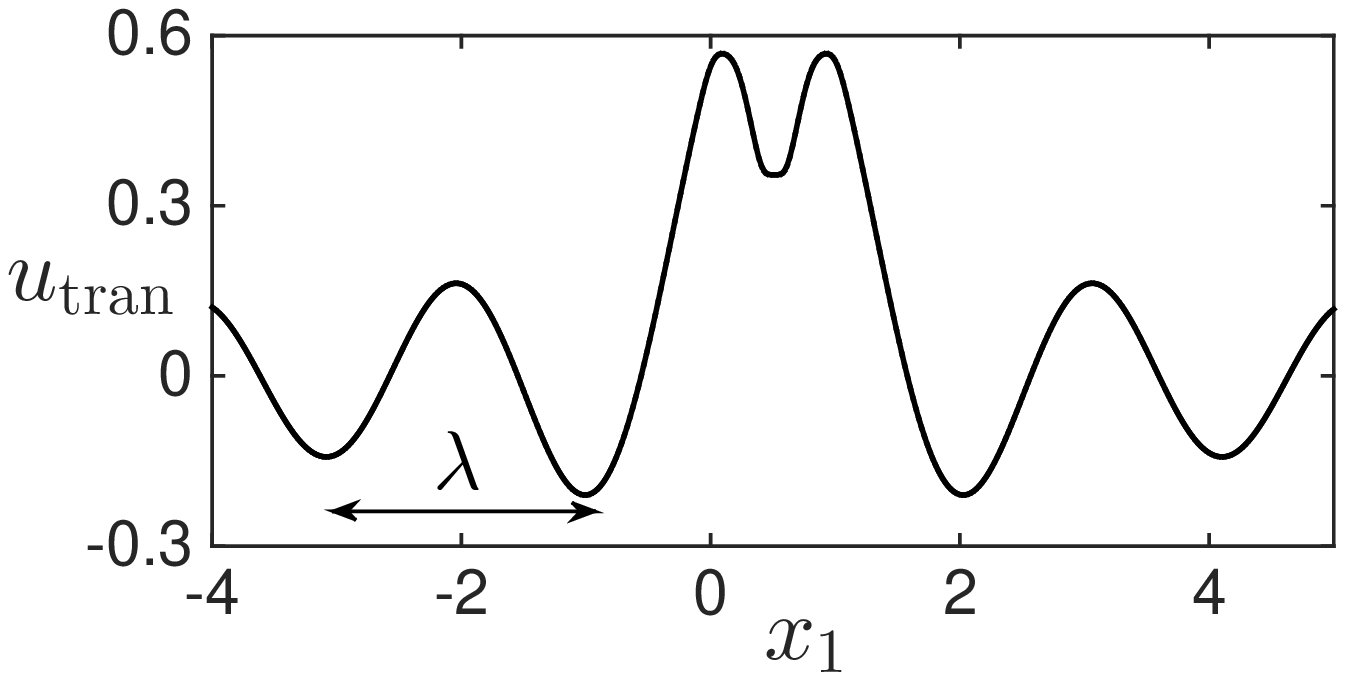} \hspace*{-0.5cm}
\includegraphics[height=3.5cm]{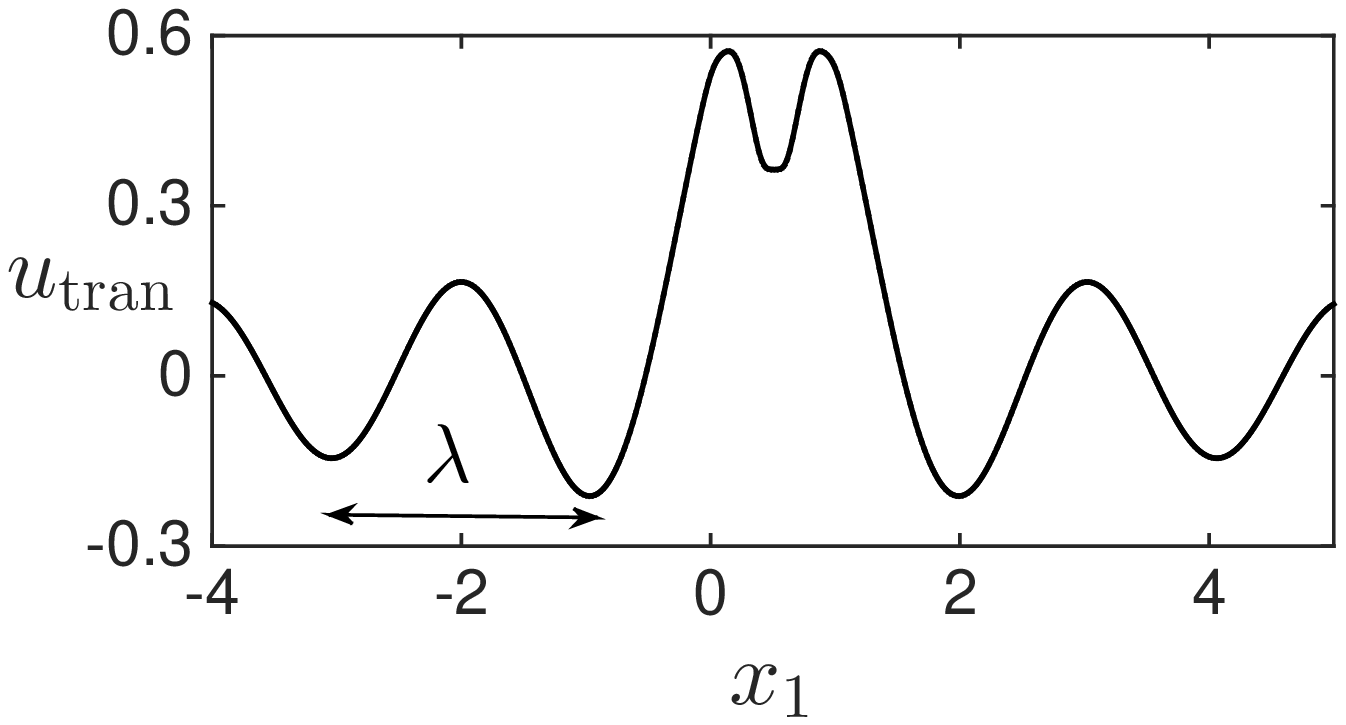} \hspace*{-0.5cm} \\
\hspace*{-0.35cm}
\includegraphics[height=3.5cm]{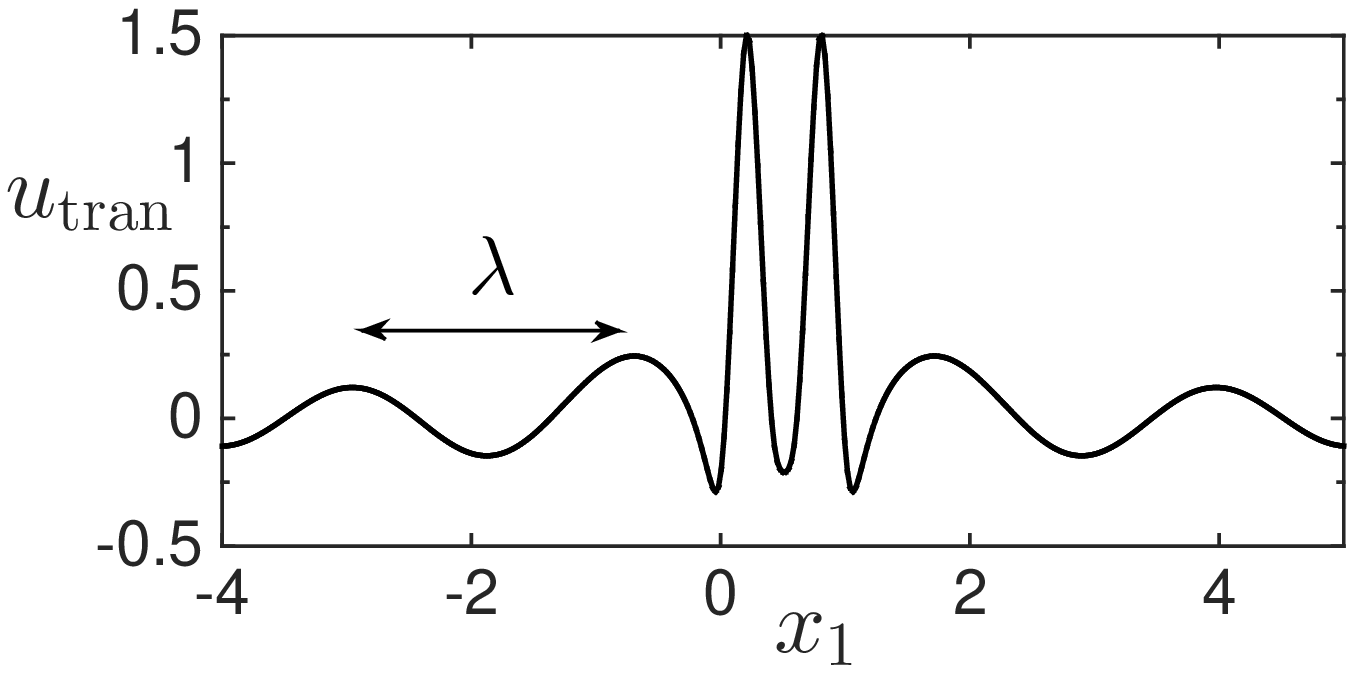} \hspace*{-0.5cm}
\includegraphics[height=3.5cm]{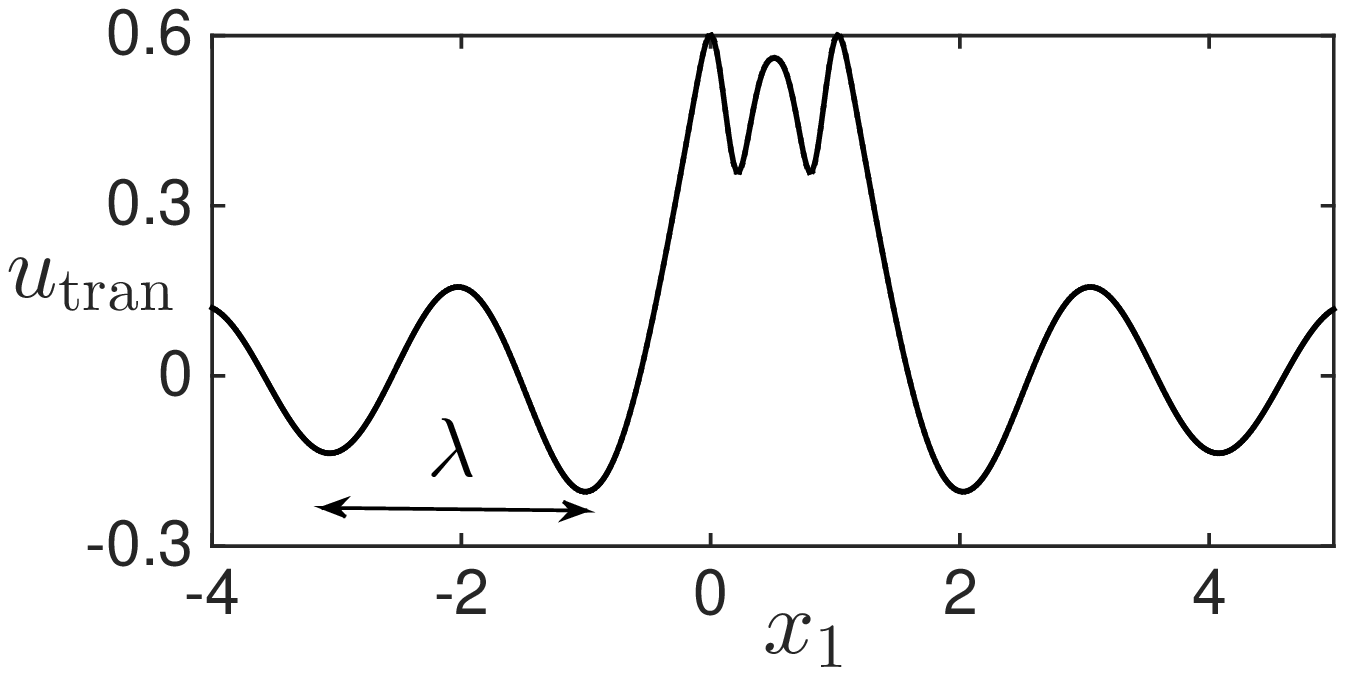} \hspace*{-0.5cm}
\caption{The real part of the transmitted field $u_\mathrm{tran}$ on the sample plane $x_2=0$ at the resonant frequencies
$k=\Re \, k_{1j}$ for $j=1$, $2$, $\cdots$, $6$. Note the the slit aperture covers the region
$I_\mathrm{slit}=[0,1]$. }\label{fig:ut1_resonance}
\end{center}
\end{figure}

\begin{figure}[!htbp]
\begin{center}
\vspace*{-0.5cm}\hspace*{-0.8cm}
\includegraphics[height=2cm]{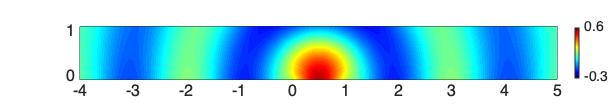}\\
\vspace*{-0.2cm}\hspace*{-0.8cm}
\includegraphics[height=2cm]{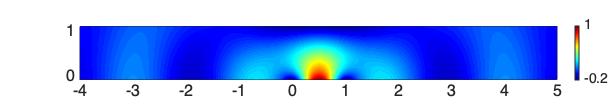} \\
\vspace*{-0.2cm}\hspace*{-0.8cm} 
\includegraphics[height=2cm]{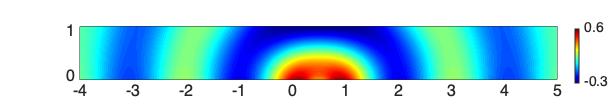} \\
\vspace*{-0.2cm}\hspace*{-0.8cm}
\includegraphics[height=2cm]{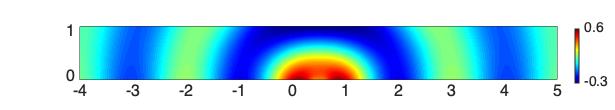} \\
\vspace*{-0.2cm}\hspace*{-0.8cm} 
\includegraphics[height=2cm]{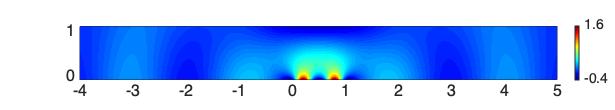} \\
\vspace*{-0.2cm}\hspace*{-0.8cm}
\includegraphics[height=2cm]{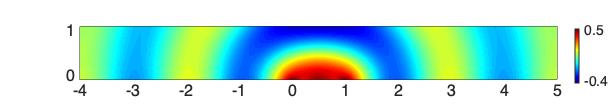}
\caption{The real part of the transmitted field $u_\mathrm{tran}$
in the region $[-4,5]\times[0,1]$ above the sample plane
at the resonant frequencies $k=\Re \, k_{1j}$ for $j=1$, $2$, $\cdots$, $6$.}\label{fig:ut2_resonance}
\vspace*{-20pt}
\end{center}
\end{figure}

\bigskip

Note that at the resonant frequency $k=\Re \, k_{mj}$,  the transmitted field $u_\mathrm{tran}$ will be amplified by an order 
that is inversely proportional to $\Im \, k_{mj}$. On the other hand, due to the smallness of $\delta$,
$u_\mathrm{tran}$ can be viewed as the field generated by an array of point charges $\alpha_j$ located at the hole aperture $\Gamma_{j}^{(2)}$ \cite{lin_zhang17, lin_zhang18_2}.
Depending the resonance frequencies, $\alpha_j$ can be positive or negative, which would induce different oscillatory patterns for $u_\mathrm{tran}$.
This is illustrated in Figures \ref{fig:ut1_resonance} and \ref{fig:ut2_resonance} for $J=6$, where the transmitted field $u_\mathrm{tran}$ 
on the sample plane $x_2=0$ and  in the region $[-4,5]\times[0,1]$ above the sample plane is shown 
at the resonant frequencies $k=\Re \, k_{1j}$ for $j=1$, $2$, $\cdots$, $6$, respectively.

\section{Super-resolution imaging of infinitely thin samples}\label{sec:inf-thin-sample}
\subsection{Formulation of the imaging problem}
We first consider the configuration where the sample is an infinitely thin sheet.
The thin sheet can be produced, for instance by microcontact printing \cite{XW}. 
This allows us to ignore the topography induced effect and the multiple scattering between the illumination and the sample \cite{NB}.

Let $u_\mathrm{tran}$ be the transmitted field through the subwavelength holes.
Assume that the thin sheet is characterized by the transmission function $q(x_1)$ with  $0\le q(x_1)\le 1$.
Then the wave field after being transmitted immediately through the sample is given by 
$u_\mathrm{samp}(x_1,0)=q(x_1) u_\mathrm{tran}(x_1,0)$.
The propagation of the sample field $u_\mathrm{samp}$ to the detection plane is described by the propagator (transfer function) in the Fourier domain:
\begin{equation}\label{eq:u_d_hat}
\hat{u}_\mathrm{det}(\xi,d) = e^{i\rho_0(\xi) d} \, \hat{u}_\mathrm{samp}(\xi,0), 
\end{equation}
where
\begin{equation*}
\rho_0(\xi)= \left\{
\begin{array}{lll}
\medskip
\sqrt{k^2\varepsilon_0-\xi^2},  & |\xi| \le k, \\
\medskip
i\sqrt{\xi^2-k^2\varepsilon_0},  & |\xi| > k.
\end{array}
\right.
\end{equation*}
This translates into the wave field in the spatial domain:
\vspace*{-2pt}
\begin{equation}\label{eq:u_d1}
\vspace*{-2pt}
u_\mathrm{det}(x_1,d) = \int_{-\infty}^\infty e^{i (\xi x_1+\rho_0(\xi) d)} \, \hat{u}_\mathrm{samp}(\xi,0) \; d\xi,
\end{equation}
or equivalently, the convolution
\vspace*{-2pt}
\begin{equation}\label{eq:u_d2}
\vspace*{-2pt}
u_\mathrm{det}(\cdot,d)  =  w_d * u_\mathrm{samp} = w_d * (q\cdot u_\mathrm{tran}),
\end{equation}
where $\hat{w}_d(\xi) = e^{i \rho_0(\xi) d}$. 
Due to the exponential decay of the propagator $e^{i\rho_0(\xi) d}$ for large $|\xi|$,
in the far field where $d \gg \lambda$,  only the plane wave components with $\xi^2+\rho_0^2 \le k^2$ will reach the detector plane.

Define $p(x_1):=1-q(x_1)$, which vanishes outside the interval $I_0$.
Let $I_d$ be the measurement aperture on the detector plane.
We define the operator $A_{k}: L^2(I_0) \to L^2(I_d)$:
$$ A_k[p] =  w_d * (p\cdot u_\mathrm{tran}). $$
Let $h_k$ and $h_k^{(0)}$ be the measurement of the field over the aperture $I_d$ when the sample is present and not, respectively.
We seek to recover $p$ by solving the equation 
\begin{equation}\label{eq:linear_eqn}
A_k[p]  + \eta_k = g_k,
\end{equation}
where $g_k=h_k-h_k^{(0)}$ and $\eta_k$ denotes the noise.
In the case of multiple frequency configuration with incident frequencies $k=k_1, k_2, \cdots, k_m$, 
we assemble all the data together to solve the equation
\begin{equation}\label{eq:linear_eqn_multi_freq}
A[p] + \eta =g,
\end{equation}
where the operator $A: L^2(I_0) \to \big(L^2(I_d)\big)^m$, the measurement $g$ and the noise $\eta$ are given by 
\begin{equation*}
A=\left
[\begin{array}{cccc}
A_{k_1}  \\
\cdot    \\
\cdot    \\
\cdot   \\
A_{k_m}
\end{array}
\right],
\quad
g=\left
[\begin{array}{cccc}
g_{k_1}  \\
\cdot    \\
\cdot    \\
\cdot   \\
g_{k_m}
\end{array}
\right],
\quad \mbox{and} \quad
\eta=\left
[\begin{array}{cccc}
\eta_{k_1}  \\
\cdot    \\
\cdot    \\
\cdot   \\
\eta_{k_m}
\end{array}
\right].
\end{equation*}

\subsection{Reconstruction algorithms}

\subsubsection{Gradient descent method}
The most natural approach for solving the equation \eqref{eq:linear_eqn_multi_freq} is to formulate it as the minimization problem
\begin{equation}\label{eq:mini_L2}
\min_{p\in L^2(I_0)} \| A[p] -g \|_{L^2(I_d)},
\end{equation}
and apply the gradient descent algorithm. By starting at $p_0=0$, the iteration is computed as follows:
$$ p_{n+1} = p_n +  \alpha_n  r_n,  \quad n \ge 0 $$
in which
$$ r_n = \Re (A^*g-A^*A p_n), \quad \alpha_n = \frac{\langle r_n, r_n\rangle}{\langle A^*A r_n, r_n \rangle}. $$

In view of \eqref{eq:u_d_hat}, when $d \gg \lambda $ the convolution operator in \eqref{eq:u_d2} is smoothing
which essentially filters the frequency components of a function outside the frequency band $\Omega_k$. 
Therefore, with the application of the operator $A^*$ at each step, the gradient descent iteration
is very insensitive to the highly oscillatory noise in the measurement.
On the other hand, as pointed out in the formula \eqref{eq:freq_shift},
the high spatial frequency components of the function $p$ outside the band $\Omega_k$ is transferred to $\Omega_k$ when 
transmitted wave fields with different oscillation patterns interact with the sample, and these frequency components can be reconstructed from the measurement $g$.
These two features together yield a super-resolution imaging of $p$ in a stable manner.  \\

\subsubsection{Total variation regularization and the split Bregman iteration}
To capture the sharp edges in the image, one can apply the total variation regularization for the reconstruction \cite{ROF}. 
This boils down to solving the minimization problem
\begin{equation}\label{eq:mini_TV1}
\min_{p}  \| p \|_{TV} + \frac{\alpha}{2}  \| A[p] -g \|_{L^2(I_d)},
\end{equation}
in which $\alpha>0$ is the relaxation parameter, and the total variation norm is defined as $ \| p \|_{TV} = \displaystyle{\int_{I_0} |p'| dx_1}$.
The minimization problem \eqref{eq:mini_TV1} is reformulated equivalently as the following constrained optimization problem:
\begin{equation}\label{eq:mini_TV2}
\min_{p}  \| s \|_{L^1(I_0)} + \frac{\alpha}{2}  \| A[p] -g \|_{L^2(I_d)} \quad s. t. \quad s=p',
\end{equation}
which can be converted to an unconstrained optimization problem with the relaxation parameter $\beta>0$:
\begin{equation}\label{eq:mini_TV3}
\min_{p, s}  \| s \|_{L^1(I_0)} + \frac{\alpha}{2}  \| A[p] -g \|_{L^2(I_d)} + \frac{\beta}{2}  \| s-p' \|_{L^2(I_0)}.
\end{equation}
The optimization problem \eqref{eq:mini_TV3} can be solved by the Bregman iteration method \cite{Osher1, Osher2}:
\begin{eqnarray*}
(p_{n+1}, s_{n+1}) &=& \mathrm{argmin}_{p, s}  \| s \|_{L^1(I_0)} + \frac{\alpha}{2}  \| A[p] -g \|_{L^2(I_d)} \\
&&  + \frac{\beta}{2}  \| s-p' -b_n\|_{L^2(I_0)}, \\
b_{n+1} &=& b_n + s_{n+1}-p'_{n+1}.
\end{eqnarray*}
Here $b$ is an auxiliary variable.

The split Bregman iteration is to split the objective functional into two components
and solve for $p_{n+1}$ and $s_{n+1}$ above in an alternative manner \cite{Osher1}.
The algorithm is described as follows: 

\begin{description}
\item  Set $p_0=0$, $s_0=0$, $b_0=0$.
\item  While $\|p_{n+1}-p_n\|_{L^2(I_0)} > tol $
\begin{description}
\item  $p_{n+1} = \mathrm{argmin}_{p}  \frac{\alpha}{2}  \| A[p] -g \|_{L^2(I_d)}  + \frac{\beta}{2}  \| s_n-p' -b_n\|_{L^2(I_0)}$,
\item  $s_{n+1} = \mathrm{argmin}_{s}  \| s \|_{L^1(I_0)} + \frac{\beta}{2}  \| s-p'_{n+1} -b_n\|_{L^2(I_0)}$,
\item  $b_{n+1} = b_n + s_{n+1}-p'_{n+1}$,
\end{description}
\item End
\end{description}
Since $p$ and $s$  are decoupled in each subproblem,
the optimization for $p$ and $s$ at each iteration can be obtained efficiently by solving a Possion equation for the former and applying a shrinkage operator for the latter \cite{Osher1}.

\subsection{Numerical examples}
A total of $9$ slit holes, each with width $\delta=0.02$, are patterned in a metallic slab of thickness $L=1$.
The distance between two adjacent slit holes is $\ell=0.5$ and the $9$ slit apertures span the interval $I_\mathrm{slit} :=[0, 4]$.
Among the $9$ resonant frequencies $\{k_{1,j}\}_{j=1}^9$ near $\pi$,
we choose $6$ frequencies to generate illumination patterns with distinct features.
For each frequency $k_{1,j}$, two illuminations are generated using the real and imaginary part of normal incident wave $e^{ik_{1,j} x_2}$ respectively.
The smallest resonant frequency is $k_1=2.8301$ and the largest resonant frequency is $k_6=2.9695$.
In the following examples, we use the wavelength corresponding to the largest resonant frequency 
(or the shortest wavelength) among $12$ illuminations in the discussion of resolution, and this corresponds to a wavelength $\lambda \approx 2$.
 
\begin{figure}[!htbp]
\begin{center}
\includegraphics[height=3.5cm]{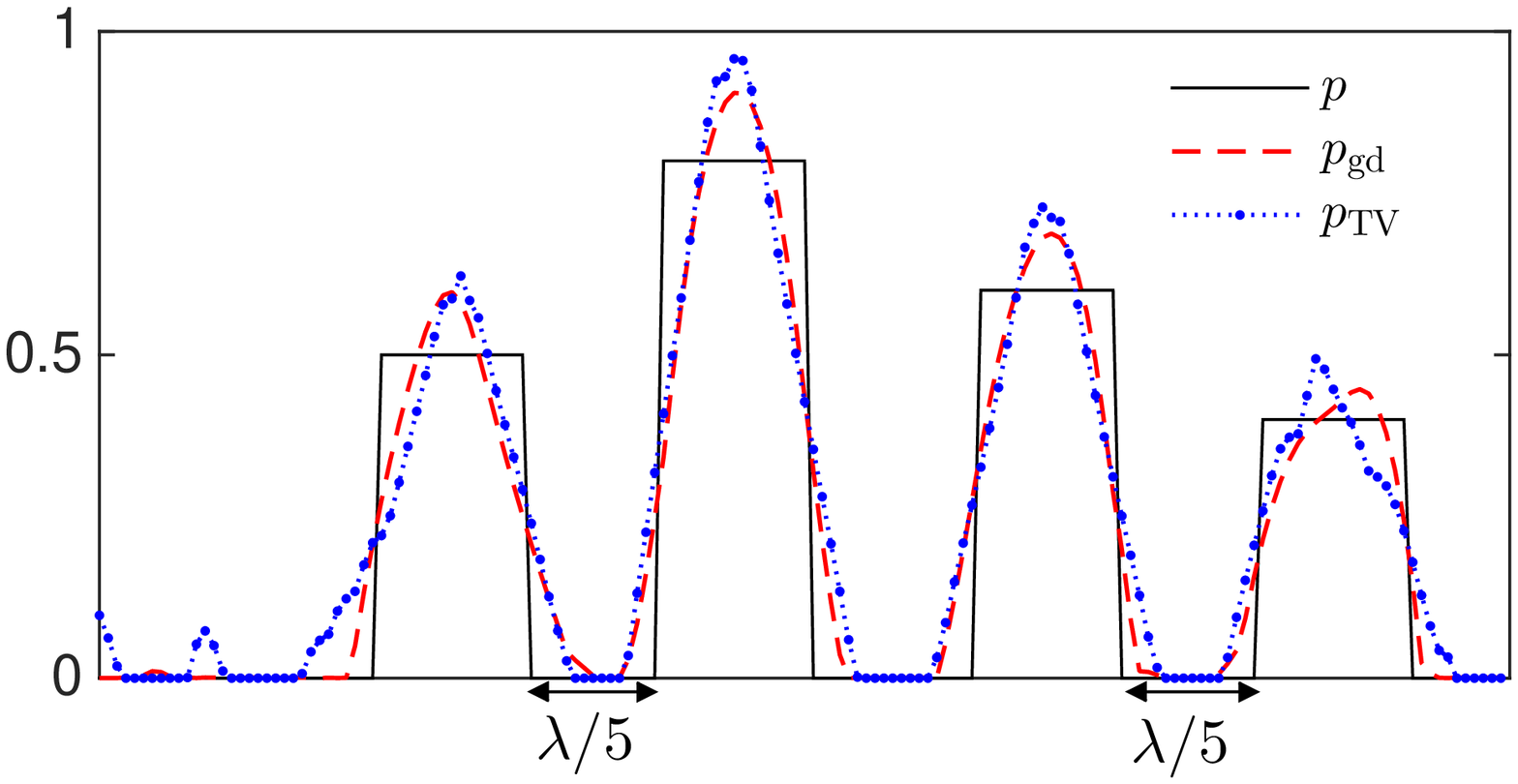} 
\includegraphics[height=3.5cm]{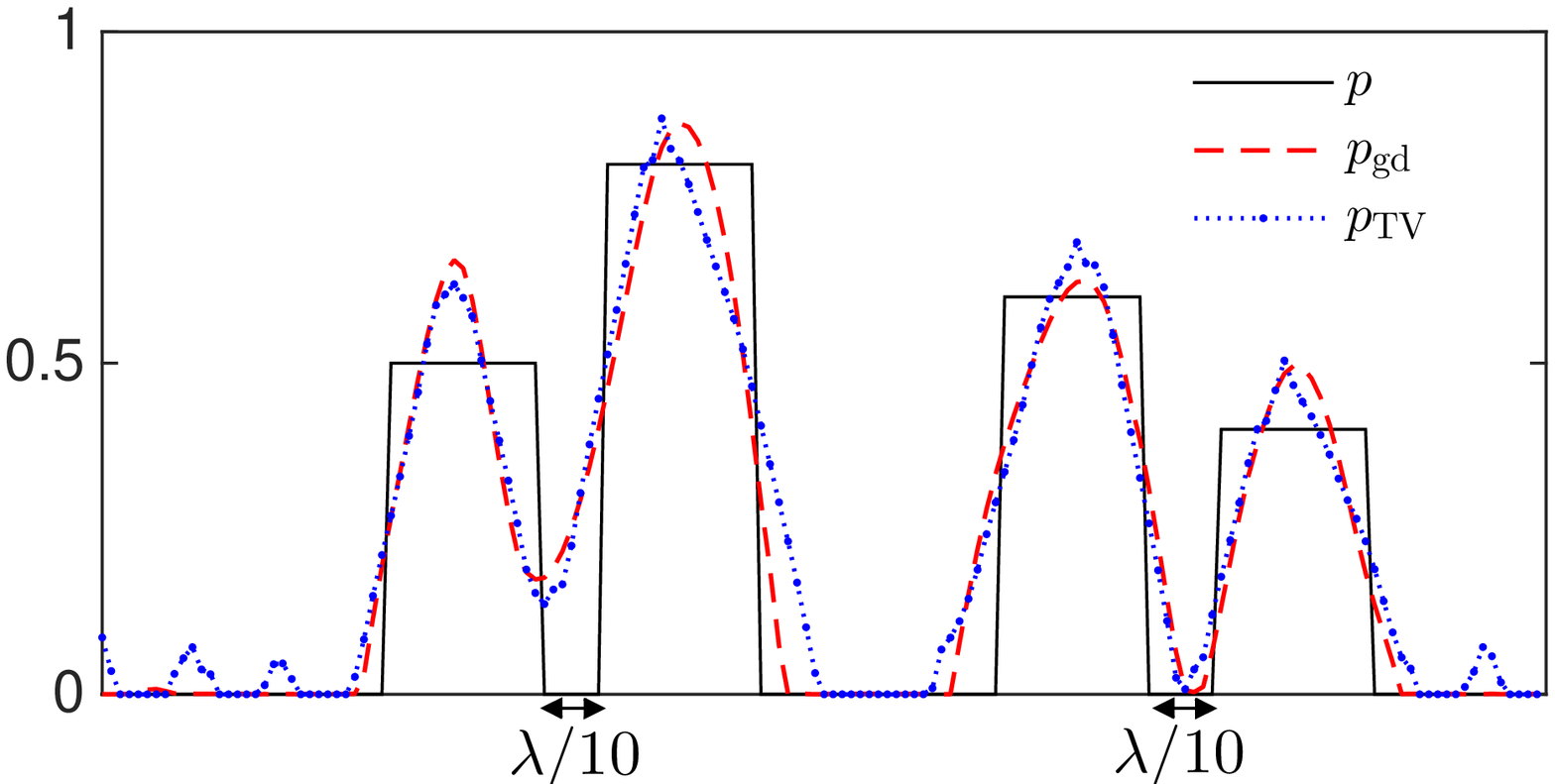} 
\caption{Reconstruction with the gradient descent method ($p_\mathrm{gd}$) and the TV regularization ($p_\mathrm{TV}$)
when $p$ is given by \eqref{p_ex1}. Left: $w=\lambda/5$; right: $w=\lambda/10$.}\label{fig:ex1}
\end{center}
\end{figure}

We first consider an infinitely thin sample with the transmission function given by $q=1-p$, where
\begin{eqnarray}\label{p_ex1}
p = 0.5 \, \rchi_{(0.8,1.2)} + 0.8 \, \rchi_{(0.8+w,1.2+w)} + 0.6 \, \rchi_{(2.5,2.9)} + 0.4 \, \rchi_{(2.5+w,2.9+w)}.
\end{eqnarray}
In the above, $\rchi_I$ denotes the characteristic function that vanishes outside the interval $I$.
We reconstruct the function over the interval $I_0=I_\mathrm{slit}$, by
setting the measurement aperture to be $I_d=[-2\lambda, 4\lambda]$ over the detector plane $x_2=5\lambda$.
Here and henceforth, $5\%$ Gaussian random noise $\eta_k$ is added to each set of synthetic data.
Figure \ref{fig:ex1} demonstrates the reconstructions when $w=\lambda/5$ and $w=\lambda/10$, respectively.
We observe that both the gradient descent algorithm and the Split Bregman iteration with TV regularization give rise to images with a resolution of $\lambda/10<\ell/2$.

\begin{figure}[!htbp]
\begin{center}
\includegraphics[height=3.5cm]{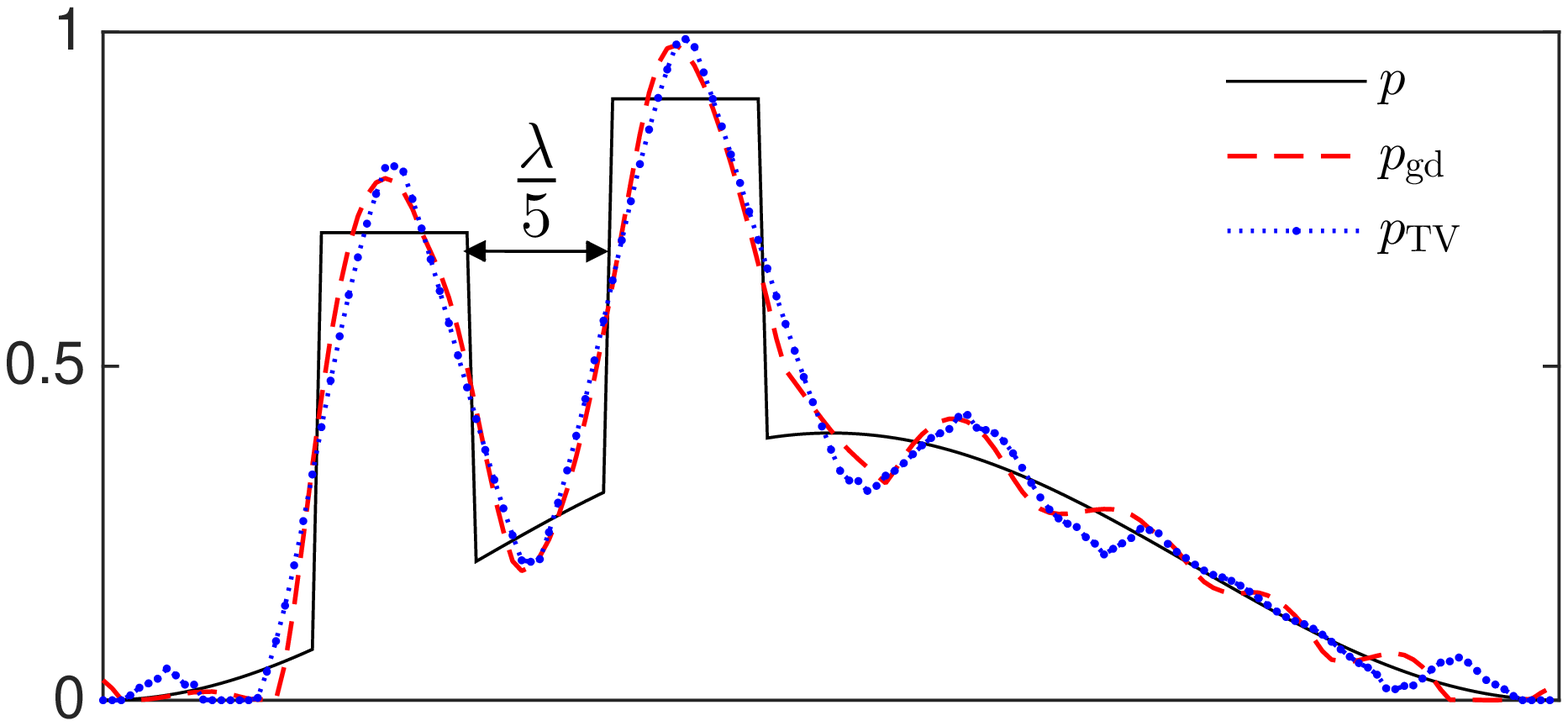} 
\includegraphics[height=3.5cm]{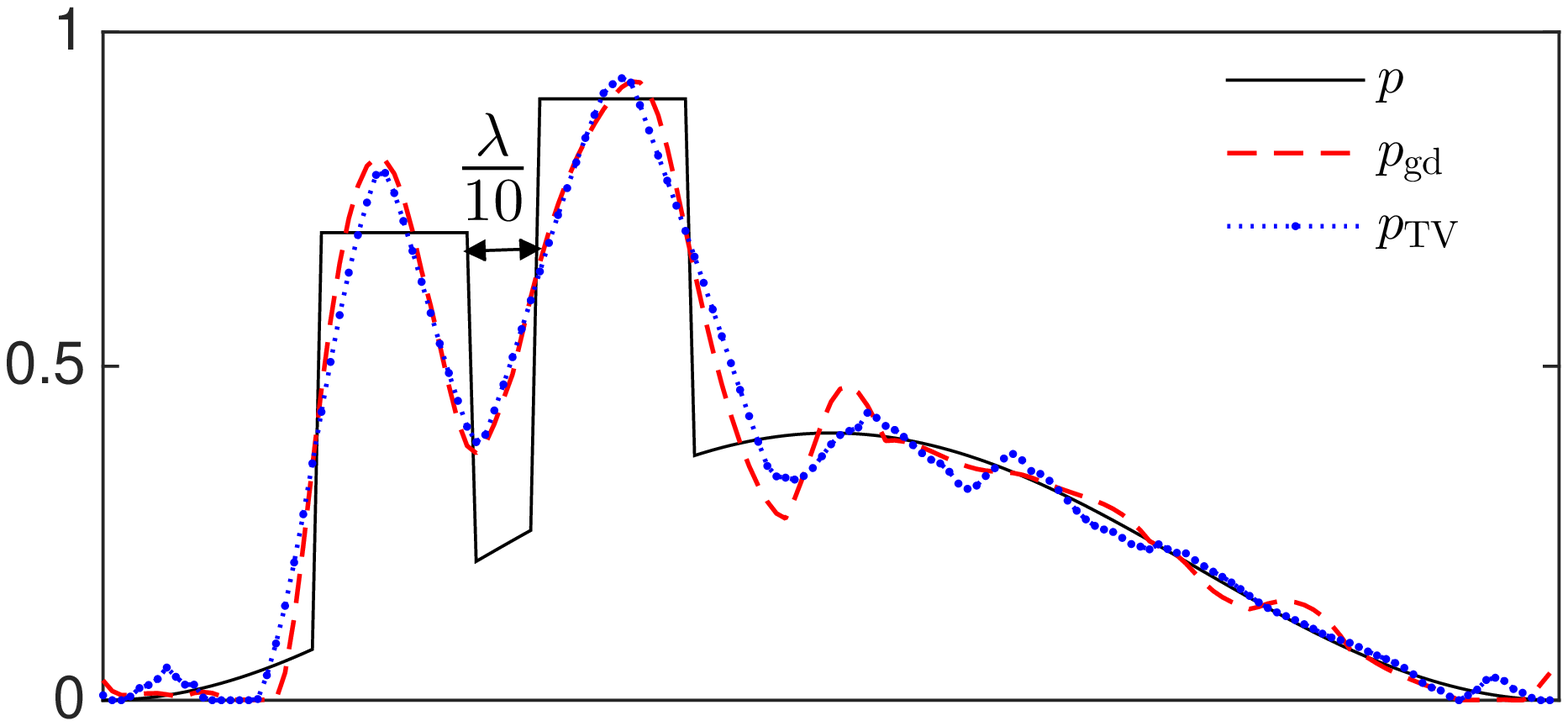} 
\caption{Reconstruction with the gradient descent method ($p_\mathrm{gd}$) and the TV regularization ($p_\mathrm{TV}$)
when $p$ is given by \eqref{p_ex2}. Left: $w=\lambda/5$; right: $w=\lambda/10$.}\label{fig:ex2}
\end{center}
\end{figure}

In the second example, we use the same subwavelength structure as above and consider the imaging of a multi-scale profile, 
where two small inhomogeneities are embedded in a smooth background.
The transmission function is expressed by $q=1-p$, in which
\begin{eqnarray}\label{p_ex2}
 p = 0.2\left(1-\cos\left(\frac{\pi}{2}x_1\right)\right)  \cdot \left( \rchi_{I_0} - \rchi_{(0.6,1)} - \rchi_{(0.6+w,1+w)} \right)
  + 0.7 \, \rchi_{(0.6,1)} + 0.9 \, \rchi_{(0.6+w,1+w)}.
\end{eqnarray}
It is seen from Figure \ref{fig:ex2} that both the inhomogeneities and the background are successfully reconstructed, 
and the image resolution remains the same as in the previous example.

\begin{figure}[!htbp]
\begin{center}
\includegraphics[height=3.5cm]{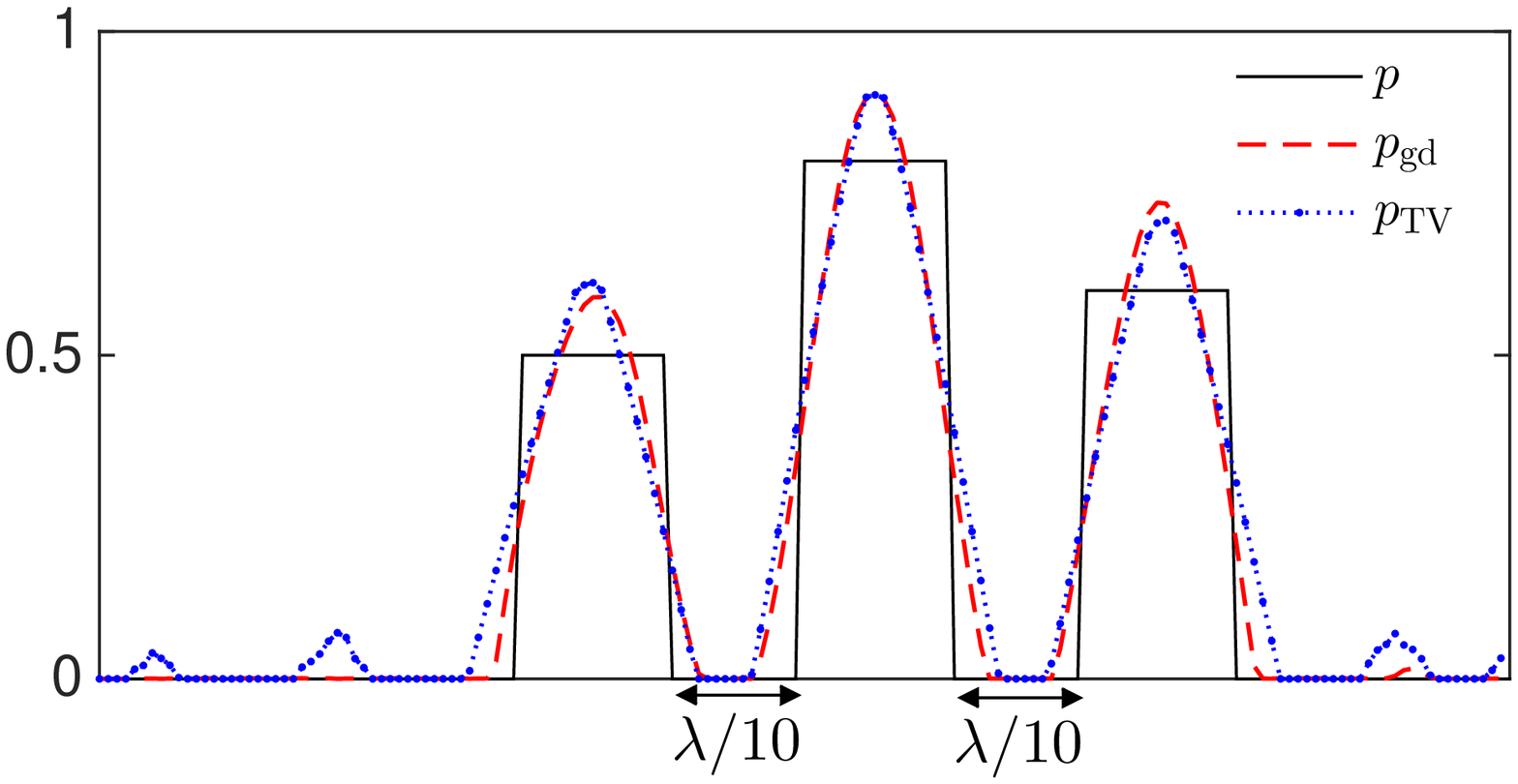} 
\includegraphics[height=3.5cm]{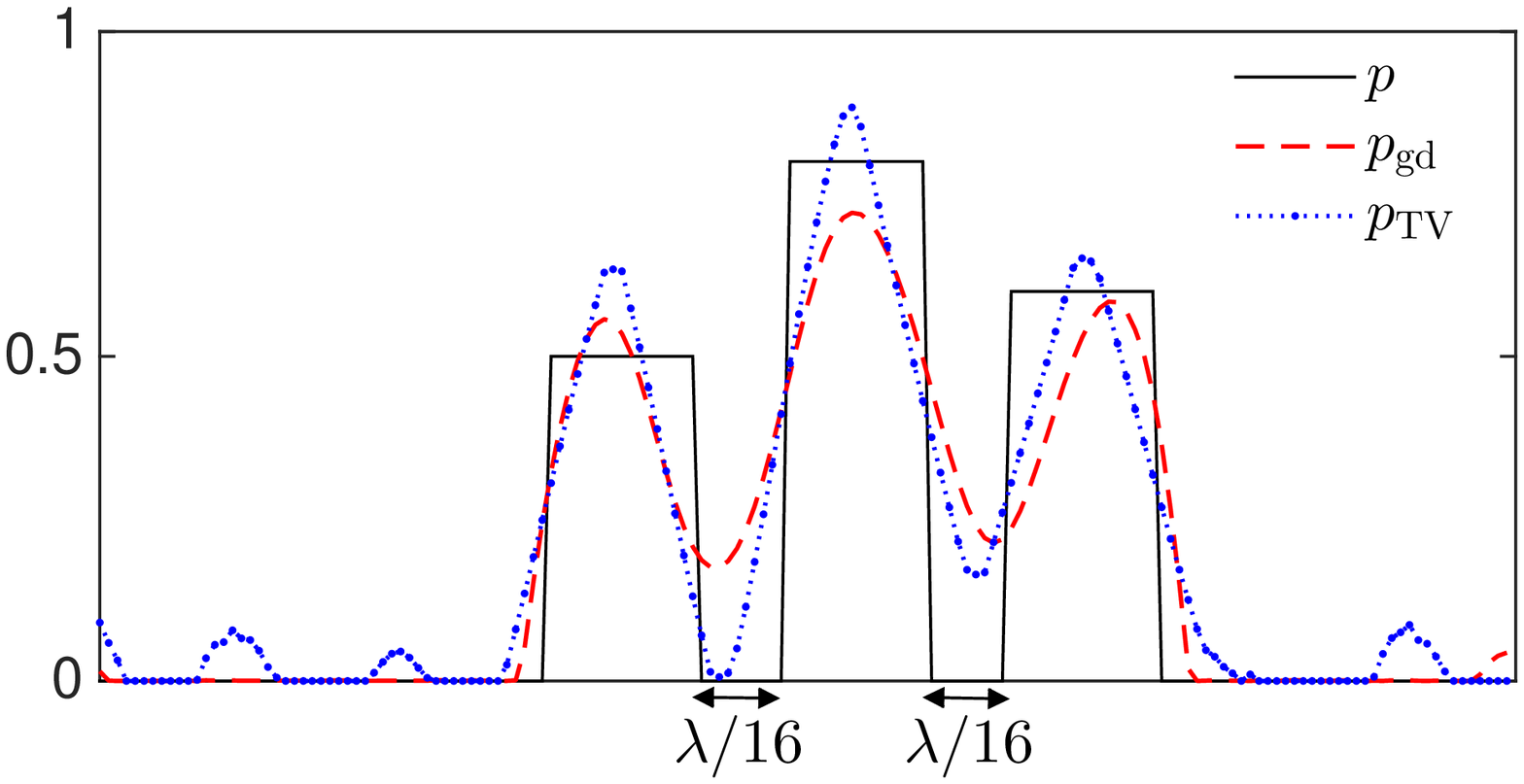} 
\caption{Reconstruction with the gradient descent method ($p_\mathrm{gd}$) and the TV regularization ($p_\mathrm{TV}$)
when $p$ is given by \eqref{p_ex3}. Left: $w=\lambda/10$; right: $w=\lambda/16$.}\label{fig:ex3}
\end{center}
\end{figure}

In the third example, we decrease the distance between the slit holes by setting $\ell=0.25$ so that
the slit apertures span the region $I_\mathrm{slit} :=[0, 2]$. Among the $6$ chosen resonant frequencies, 
the smallest frequency is $k_1=2.7877$ and the largest frequency is $k_6=2.9908$. The latter corresponds to a wavelength $\lambda \approx 2$.

Let us consider an infinitely thin sample where
\begin{eqnarray}\label{p_ex3}
 p= 0.5 \, \rchi_{(0.8-0.5w,1-0.5w)} + 0.8 \, \rchi_{(1+0.5w,1.2+0.5w)}
   + 0.6 \, \rchi_{(1.2+1.5w,1.4+1.5w)}.
\end{eqnarray}
The reconstructions are performed over the interval $I_0=I_\mathrm{slit}$,
using the measurement aperture $I_d=[-2\lambda, 3\lambda]$ over the detector plane $x_2=5\lambda$.
The images obtained from the two approaches in Figure \ref{fig:ex3}  
show that a resolution of $\lambda/16 \approx \ell/2$ can be achieved.
We also observe that in this configuration, the image obtained by the TV regularization has sharper resolution than the one obtained by the gradient descent algorithm.
It should be pointed out that one needs adjust the relaxation parameters $\alpha$ and $\beta$ in the TV optimization problem \eqref{eq:mini_TV3} carefully
to gain a sharper resolution. The optimal choice of parameters is a delicate and interesting question, and may deserve further investigation.

\section{Super-resolution imaging of thin samples with finite thickness}\label{sec:thin-sample}
In this section, we consider the configuration where a sample of finite thickness occupies the domain $R_0\subset D^{(3)}$ above the sample plane.
The linear imaging problem is investigated by assuming that the sample is a weak scatterer and using the Born approximation.
Let $u_\mathrm{tran}$ be the transmitted field through the slit holes at the absence of the imaging sample,
then the Helmholtz equation for the diffracted field satisfies
$$ \Delta u_{\mathrm{diff}}  + k^2 u_{\mathrm{diff}}  = -k^2 \, p \, u_{\mathrm{tran}}  \quad \mbox{in} \; D^{(3)}, $$
where $p(x)=\varepsilon(x)-1$ and the permittivity value $\varepsilon(x)=1$ outside the region $R_0$.
Using the layered Green's function $g^{(3)}(x, y)$ in the domain $D^{(2)}\cup D^{(3)}$ with the Neumann boundary condition along the metallic slab boundary
(see the appendix), the diffracted field in $D^{(3)}$ can be expressed as
\begin{eqnarray*}
u_{\mathrm{diff}} (x)  &=&  - k^2 \int_{R_0} g^{(3)}(x, y)  u_\mathrm{tran}(y) \, p(y) dy  +  \displaystyle{\sum_{j=1}^J\int_{\Gamma_{j}^{(2)}}}  g^{(3)}(x,y) (y) \, \partial_{y_2} u_{\mathrm{diff}}(y)   ds_y \\
 &\approx &  - k^2 \int_{R_0} g^{(3)}(x, y)  u_\mathrm{tran}(y) \, p(y) dy,
\end{eqnarray*}
where we have neglected the field arising from the induced current over the slit apertures $\Gamma_{j}^{(2)}$ by noting that $\delta \ll 1$.

With the abuse of notations, we still denote forward operator from the imaging sample $p$ to the diffracted field on the detector plane
by $A_k$, which is given by
$$ A_k[p] = - k^2 \int_{R_0} g^{(3)}(x_1,d; y)  u_\mathrm{tran}(y) \, p(y) dy. $$
$g_k:=u_{\mathrm{diff}} + \eta_k $ is the measurement over the aperture $I_d$, in which $\eta_k$ denotes the noise.
The reconstruction is performed in the region $R_0$ by applying both the gradient descent algorithm and the total variation regularization as discussed in Section \ref{sec:inf-thin-sample}.
Note that the total variation norm in \eqref{eq:mini_TV1}-\eqref{eq:mini_TV3} is now defined as $\| p \|_{TV} = \int_{R_0} |\nabla p| dx$,
and the derivative in the split Bregman iteration is replaced by the gradient $\nabla$.
We use the same subwavelength structures as in Section \ref{sec:inf-thin-sample} to generate of 12 illumination patterns and the numerical results are discussed below.

\begin{figure}[!htbp]
\begin{center}
\includegraphics[height=2cm]{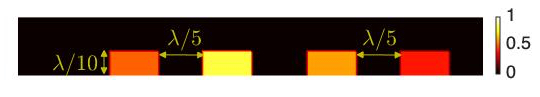}  \\
\includegraphics[height=2cm]{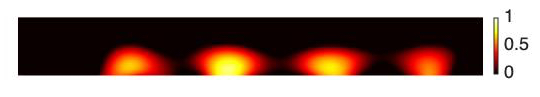}  \\
\includegraphics[height=2cm]{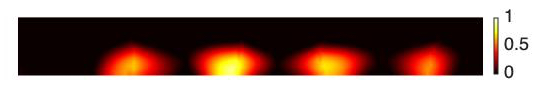}
\caption{Real image in the domain $R_0=[0,4]\times[0, 0.5]$ (top), and reconstructions with the gradient descent method (middle) and the TV regularization (bottom).
The measurement aperture is $[-2\lambda, 4\lambda]$.}\label{fig:ex1-1_thick}
\vspace*{-10pt}
\end{center}
\end{figure}

\begin{figure}[!htbp]
\begin{center}
\includegraphics[height=2cm]{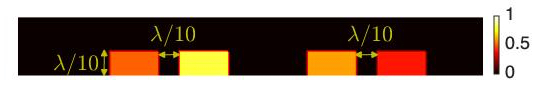}  \\
\includegraphics[height=2cm]{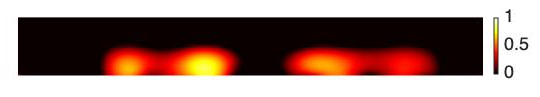}  \\
\includegraphics[height=2cm]{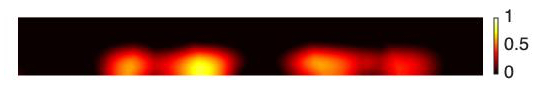}
\caption{Real image in the domain $R_0=[0,4]\times[0, 0.5]$ (top), and reconstructions with the gradient descent method (middle) and the TV regularization (bottom).
The numerical setup as in Figure \ref{fig:ex1-1_thick} is used. }\label{fig:ex1-2_thick}
\vspace*{-10pt}
\end{center}
\end{figure}

\begin{figure}[!htbp]
\begin{center}
\includegraphics[height=2cm]{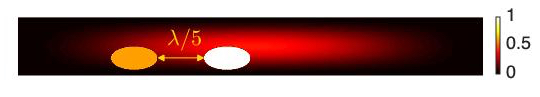}  \\
\includegraphics[height=2cm]{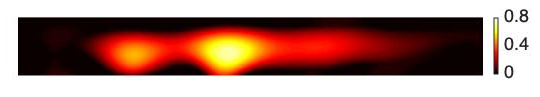}  \\
\includegraphics[height=2cm]{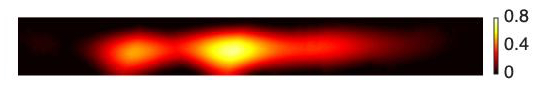}
\caption{Real image in the domain $R_0=[0,4]\times[0, 0.5]$ (top), and reconstructions with the gradient descent method (middle) and the TV regularization (bottom).
The measurement aperture is $[-2\lambda, 4\lambda]$.}\label{fig:ex2-1_thick}
\vspace*{-10pt}
\end{center}
\end{figure}

\begin{figure}[!htbp]
\begin{center}
\includegraphics[height=2cm]{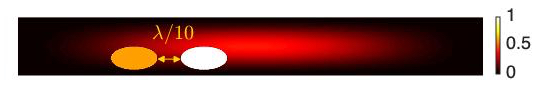}  \\
\includegraphics[height=2cm]{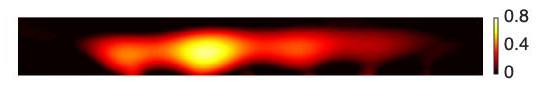}  \\
\includegraphics[height=2cm]{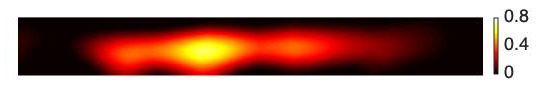}
\caption{Real image in the domain $R_0=[0,4]\times[0, 0.5]$ (top), and reconstructions with the gradient descent method (middle) and the TV regularization (bottom).
The numerical setup as in Figure \ref{fig:ex2-1_thick} is used.}\label{fig:ex2-2_thick}
\vspace*{-10pt}
\end{center}
\end{figure}

When the distance between the adjacent slit holes is $\ell=0.5$ and the $9$ slit apertures span the interval $[0, 4]$, 
we set the reconstruction domain as $R_0=[0,4]\times[0, 0.5]$ and the detector plane is placed over $x_2=5\lambda$,
where $\lambda$ again denotes the wavelength corresponding to the largest resonant frequency.
Figures \ref{fig:ex1-1_thick} and \ref{fig:ex1-2_thick} show the real image and the reconstructions when the sample consists of 
four rectangular shape scatterers. The measurement aperture is $I_d=[-2\lambda, 4\lambda]$ so that
 the aperture size is $5\lambda$. It is seen that a resolution of $\lambda/10<\ell/2$ is achieved in both numerical reconstructions.
The same subwavelength structure is used to illuminate the sample that consists of two oval type scatterers embedded in an inhomogeneous background medium,
and a resolution $\lambda/5$ is obtained (see Figure \ref{fig:ex2-1_thick}).
The image quality deteriorates in this scenario as the two scatterers get closer. This is shown in Figure \ref{fig:ex2-2_thick}, where the distance between two oval scatterers is $\lambda/10$
for the real image. The loss of accuracy is due to scattering induced by the inhomogeneous background medium.

\begin{figure}[!htbp]
\begin{center}
\hspace*{-1cm}
\includegraphics[height=3cm]{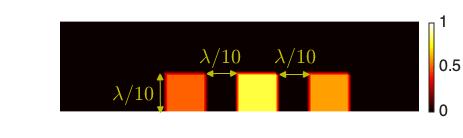}  \\
\hspace*{-1cm}
\includegraphics[height=3cm]{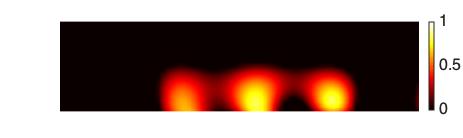}   \\
\hspace*{-1cm}
\includegraphics[height=3cm]{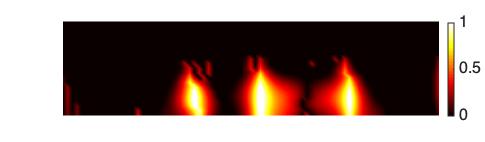}  
\caption{Real image in the domain $R_0=[0,2]\times[0, 0.5]$ (top), and reconstructions with the gradient descent method (middle) and the TV regularization (bottom).
The measurement aperture is $[-2\lambda, 3\lambda]$. }\label{fig:ex3-1_thick}
\vspace*{-10pt}
\end{center}
\end{figure}

\begin{figure}[!htbp]
\begin{center}
\hspace*{-1cm}
\includegraphics[height=3cm]{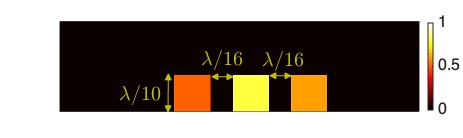} \\
\hspace*{-1cm}
\includegraphics[height=3cm]{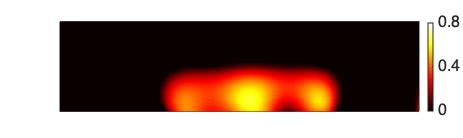} \\
\hspace*{-1cm}
\includegraphics[height=3cm]{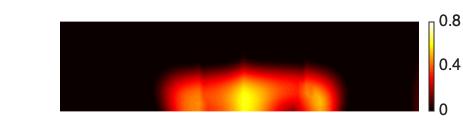} 
\caption{Real image in the domain $R_0=[0,2]\times[0, 0.5]$ (top), and reconstructions with the gradient descent method (middle) and the TV regularization (bottom).
The measurement aperture is $[-2\lambda, 3\lambda]$.}\label{fig:ex3-2_thick}
\vspace*{-10pt}
\end{center}
\end{figure}

\begin{figure}[!htbp]
\begin{center}
\hspace*{-1cm}
\includegraphics[height=3cm]{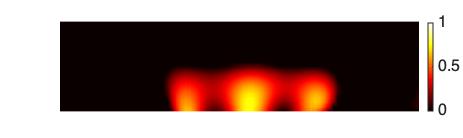} \\
\hspace*{-1cm}
\includegraphics[height=3cm]{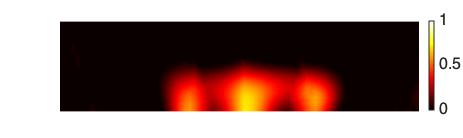} 
\caption{Reconstructions of the sample in Figure \ref{fig:ex3-2_thick} with the measurement aperture $[-4\lambda, 5\lambda]$.
Top: the gradient descent method; bottom: TV regularization.}\label{fig:ex3-3_thick}
\vspace*{-10pt}
\end{center}
\end{figure}

We next decrease the distance between the slit holes to  $\ell=0.25$ so that
the slit apertures span the interval $[0, 2]$. The reconstruction is performed in the domain $R_0=[0,2]\times[0, 0.5]$.
Figures \ref{fig:ex3-1_thick} and \ref{fig:ex3-2_thick} depict the reconstructed images with a resolution of $\lambda/10$  and $\lambda/16 \,(\approx \ell/2)$ respectively,
where the measurement aperture is also set as $I_d=[-2\lambda, 3\lambda]$ in the numerical simulation.
If one increases the measurement aperture size, then the boundary between the two close scatterers becomes clearer.
This is demonstrated in Figure \ref{fig:ex3-3_thick}, where the sample in Figure \ref{fig:ex3-2_thick} is reconstructed using the measurement over 
a larger interval $[-4\lambda, 5\lambda]$.
We also point out that, in all numerical examples, the images obtained by the TV regularization attain sharper edges compared with the gradient descent algorithm.

\section{Conclusion and discussion}
In this paper, we have presented a super-resolution imaging approach by using subwavelength hole resonances.
An array of resonant holes are arranged close to each other to generate illumination patterns that can probe
both the low and the high spatial frequency components of the imaging sample so as to break the diffraction limit.
Numerical approaches based on the gradient descent and total variation regularizations were developed to 
perform the reconstruction from the far-field measurement.
It is shown that the resolution of the reconstructed images is determined by the distance between the subwavelength holes.
Furthermore, the numerical reconstruction is stable against noise.

The ongoing studies for the three-dimensional problem will be reported elsewhere, where one can use annular subwavelength holes
to generated the desired illumination patterns to probe the sample. Another interesting question is to investigate the fully nonlinear inverse problem 
when multiple scattering between the illumination wave and the imaging sample is significant and the Born approximation fails.

Finally, we would like to point out that the proposed imaging approach allows for a very high lateral resolution in the sample plane. But it could not improve the axial resolution. 
This is because the highly oscillatory transmitted field through the subwavelength structure are localized near the slit holes due to their evanescent nature
 (see Figure \ref{fig:ut2_resonance}), and they can not propagate very deep to probe the sample in the axial direction. The improvement of the resolution
 in the axial direction is a very challenging problem and deserves further research efforts.

\appendices
\section{Green's functions in the layered medium}\label{appendixA}
We derive the Green's function in the layered medium $D^{(2)}\cup D^{(3)}$ 
with the Neumann boundary condition on the bottom of $D^{(2)}$.
Recall that  $D^{(2)}$ and $D^{(3)}$ denotes the substrate domain and the domain above the substrate respectively (see Figure \ref {fig:slits_imaging}).
The Green's function $g^{(3)}(x,y)$ for $y\in D^{(3)}$ satisfies
\begin{align*}
&\Delta_x g^{(3)}(x,y) + k^2 \varepsilon(x) g^{(3)}(x,y) = \delta_y  \quad \mbox{in} \; D^{(2)}\cup D^{(3)}, \\
&[g^{(3)}(x,y)]=\left[ \frac{1}{\varepsilon} \partial_2 g^{(3)}(x,y) \right] =0 \quad \; \mbox{for} \; x_2 = 0, \\
&\partial_2 g^{(3)}(x,y)  = 0  \quad \; \mbox{for} \; x_2 = -h. 
\end{align*}
By taking the Fourier transform of  the above equation with respect to the variable $x_1-y_1$, the Green's function in the Fourier domain solves
\begin{align*}
&\widehat{g^{(3)}}''(\xi; x_2, y_2) + (k^2 \varepsilon - \xi^2) \widehat{g^{(3)}}(\xi; x_2, y_2) = \delta_{y_2},  \; x_2>-h,  \\
&[\widehat{g^{(3)}}(\xi;  0,y_2)]=\left[ \frac{1}{\varepsilon}  \widehat{g^{(3)}}'(\xi; 0,y_2) \right] =0, \\
& \widehat{g^{(3)}}'(\xi; -h,y_2)=0.
\end{align*}
Define
\begin{eqnarray*}
\rho_0(\xi) &=& \sqrt{k^2\varepsilon_0-\xi^2}, \quad \rho_h(\xi) = \sqrt{k^2\varepsilon_h-\xi^2}, \\
w_+(\xi) &=& \rho_0 \varepsilon_h (e^{i2\rho_h h}+1) + \rho_h \varepsilon_0 (e^{i2\rho_h h}-1), \\
w_-(\xi) &=& \rho_0 \varepsilon_h (e^{i2\rho_h h}+1) - \rho_h \varepsilon_0 (e^{i2\rho_h h}-1).
\end{eqnarray*}
Then the solution of the above equation is given by
\begin{equation*}
\widehat{g^{(3)}}= \left\{
\begin{array}{llll}
\vspace*{0.1cm}
 \dfrac{1}{2i\rho_0}  e^{i\rho_0 |x_2-y_2|}  +    R \, e^{i\rho_0 (x_2+y_2)}  &&    x_2>0, \\
\vspace*{0.1cm}
T \, \cos (\rho_h(x_2+h)) e^{i\rho_0 y_2}    &&   -h<x_2<0, 
\end{array}
\right.
\end{equation*}
where the coefficients
$$ R =  \frac{w_+}{2i\rho_0 w_-}, \quad  T = \frac{2\varepsilon_h e^{i \rho_h h}  }{i w_-}.  $$
If one decomposes the coefficient $R$ as $R:= R_0 + R_1$, in which $R_0 =  \dfrac{1}{2i\rho_0}  \cdot \dfrac{\varepsilon_h-\varepsilon_0}{\varepsilon_h+\varepsilon_0}$,
then $R_1 = O(1/|\xi|^3)$ as $|\xi| \to \infty$. On the other hand, there holds $T = O(e^{-|\xi|h}/|\xi|)$  as $|\xi| \to \infty$.

Let  $H_0^{(1)}$ be the first type Hankel function of zero order.  By applying the inverse Fourier transform for $\widehat{g^{(3)}}(\xi; x_2, y_2)$ and use the identity
$$ \dfrac{1}{2\pi}\int_{-\infty}^\infty \dfrac{1}{2i\rho_0(\xi)} e^{i\rho_0(\xi)  |x_2-y_2|} e^{i\xi(x_1-y_1)} d\xi = -\dfrac{i}{4} H_0^{(1)} (k|x-y|),  $$
we obtain
\begin{equation*}
g^{(3)}(x,y) = \left\{
\begin{array}{llll}
\medskip
 -\dfrac{i}{4}\left(H_0^{(1)}(k|x-y|) + \dfrac{\varepsilon_h-\varepsilon_0}{\varepsilon_h+\varepsilon_0} H_0^{(1)}(k|x'-y|)\right)  + g^{(3)}_R(x,y),  \quad   x\in D^{(3)}, \\
 g^{(3)}_T(x,y),  \quad x\in D^{(2)},
\end{array}
\right.
\end{equation*}
where $x'$ is the reflection of $x$ by the $x_1$-axis. 
The functions $g^{(3)}_R(x,y)$ and $g^{(3)}_T(x,y)$ are the Sommerfeld integrals given by
\begin{align*}
& g^{(3)}_R(x,y) = \dfrac{1}{2\pi}\int_{-\infty}^\infty  R_1  e^{i\rho_0 (x_2+y_2)}  e^{i\xi(x_1-y_1)} \, d\xi \\
& g^{(3)}_T(x,y) = \dfrac{1}{2\pi}\int_{-\infty}^\infty  T \cos (\rho_h(x_2+h)) e^{i\rho_0 y_2}  e^{i\xi(x_1-y_1)} \, d\xi.
\end{align*}

Following similar calculations as above, it can be shown that the Green's function $g^{(2)}(x,y)$ for $y\in D^{(2)}$
takes the following form:
\begin{equation*}
g^{(2)}(x,y) = \left\{
\begin{array}{llll}
 g^{(2)}_T(x,y),  \quad x\in D^{(3)}, \\
 \medskip
 -\dfrac{i}{4}\left(H_0^{(1)}(kn_h|x-y|) + H_0^{(1)}(kn_h|x'-y|)\right)  + g^{(2)}_R(x,y),  \quad   x\in D^{(2)}.
\end{array}
\right.
\end{equation*}
In the above, $x'$ denotes the reflection of $x$ by the line $x_2=-h$. 
The corresponding Sommerfeld integrals are
\begin{align*}
 g^{(2)}_R(x,y) =  \dfrac{1}{2\pi}\int_{-\infty}^\infty  & \tilde R \, \big[ \cos (\rho_h(x_2-y_2)) +  \cos (\rho_h(x_2+y_2+2h))\big]  \, e^{i\xi(x_1-y_1)} \, d\xi, \\
 g^{(2)}_T(x,y) =  \dfrac{1}{2\pi}\int_{-\infty}^\infty & \tilde T \cos (\rho_h(y_2+h)) e^{i\rho_0x_2}  e^{i\xi(x_1-y_1)}  \, d\xi,
\end{align*}
where
$$ \tilde R = \frac{(\rho_h\varepsilon_0-\rho_0\varepsilon_h) e^{i 2\rho_h h}  }{i\rho_0 \tilde w}, \quad \tilde T = \frac{2 e^{i \rho_h h}  }{i \tilde w}  $$
and 
$$\tilde w=(\rho_h\varepsilon_0+\rho_0\varepsilon_h) - (\rho_h\varepsilon_0-\rho_0\varepsilon_h) e^{i 2\rho_h h}.$$


\begin{thebibliography}{99}

\bibitem{Abbe}
E. Abbe, {\em Beitr\"age zur Theorie des Mikroskops und der mikroskopischen Wahrnehmung}, Archiv f\"ur mikroskopische Anatomie, \textbf{9} (1873), 413-418.

\bibitem{Habib}
H. Ammari, H. Kang, B. Fitzpatrick, M. Ruiz, S. Yu and H. Zhang,
{\em Mathematical and Computational Methods in Photonics and Phononics},
Mathematical Surveys and Monographs, Volume 235, American Mathematical Society, Providence,
2018.

\bibitem{Habib-Zhang}
H. Ammari, H. Zhang, {\em  A mathematical theory of super-resolution by using a system of sub-wavelength Helmholtz resonators}, 
Commun. Math. Phys., \textbf{337} (2015), 379-428. 

\bibitem{BLi2}
G. Bao and P. Li,  {\em Near-field imaging of infinite rough surfaces},  SIAM J. Appl. Math.,  \textbf{73}  (2013), 2162-2187.

\bibitem{BLi3}
G. Bao and P. Li,  {\em Near-field imaging of infinite rough surfaces in dielectric media},  SIAM J.  Imag. Sci., \textbf{7} (2014), 867-899.

\bibitem{BL2}
G. Bao and J. Lin, {\em Near-field imaging of the surface displacement on an infinite ground plane}, Inverse Probl. Imag., \textbf{2} (2013), 377-396.

\bibitem{BL3}
G. Bao and J. Lin, {\em Imaging of reflective surfaces by near-field optics}, Opt. Lett., \textbf{37} (2012), 5027-5029.

\bibitem{Betzig}
E. Betzig \textit{et al.}, {\em Imaging intracellular fluorescent proteins at nanometer resolution}, Science, \textbf{313}  (2006), 1642-1645.

\bibitem{Blanchard-Meunier}
A. Blanchard-Dionne and M. Meunier, {\em Sensing with periodic nanohole arrays}, Advances in Optics and Photonics, \textbf{9} (2017), 891-940.

\bibitem{Candes}
E. Cand\`es and C. Fernandez-Granda, {\em Towards a mathematical theory of super-resolution}, Commun. Pure Appl. Math., \textbf{67} (2014), 906-956.

\bibitem{CMS}
P. Carney, M. Vadim, and J. Schotland, {\em Near-field tomography without phase retrieval}, Phy. Rev. Lett., \textbf{86}, (2001), 5874.

\bibitem{CS1}
P. Carney and J. Schotland, {\em Inverse scattering for near-field microscopy}, Appl. Phys. Lett., \textbf{77} (2000), 2798-800.

\bibitem{CS2}
P. Carney and J. Schotland, {\em Three-dimensional total internal reflection microscopy}, Opt. Lett., \textbf{26} (2001), 1072-1074.

\bibitem{Chen}
X. Chen, {\em Computational Methods for Electromagnetic Inverse Scattering}, Wiley-IEEE, 2018.

\bibitem{CB}
D. Courjon and C. Bainier, {\em Near field microscopy and near field optics}, Rep. Prog. Phys., \textbf{57} (1994), 989-1028.

\bibitem{CSS}
D. Courjon, K. Sarayeddine, and M. Spajer, {\em Scanning tunneling optical microsocpy}, Opt. Commun., \textbf{71} (1989), 23-28.

\bibitem{Demanet}
L. Demanet, D. Needell, and N. Nguyen, {\em Super-resolution via superset selection and pruning}, arXiv preprint arXiv:1302.6288 (2013).

\bibitem{Donoho}
D. Donoho, {\em Superresolution via sparsity constraints}, SIAM J. Math. Anal., \textbf{23} (1992), 1309-1331.

\bibitem{Dunn}
R. Dunn, {\em Near-Field Scanning Optical Microscopy}, Chem. Rev., \textbf{99} (1999), 2891-2927.

\bibitem{ELGTW}
T. Ebbesen, \textit{et al.}, {\em Extraordinary optical transmission through sub-wavelength hole arrays}, Nature, \textbf{391} (1998), 667-669.

\bibitem{Osher1}
T. Goldstein and Stanley Osher, {\em The split Bregman method for L1-regularized problems},  SIAM J Imaging Sciences, \textbf{2} (2009): 323-343.

\bibitem{Gustafasson1}
M. Gustafsson,  {\em Surpassing the lateral resolution limit by a factor of two using structured illumination microscopy}, J. of Microscopy, \textbf{198} (2000), 82-87.

\bibitem{Gustafasson2}
M. Gustafsson, {\em Nonlinear structured-illumination microscopy: wide-field fluorescence imaging with theoretically unlimited resolution}, 
 Proc. Nat. Acad. Sci., \textbf{102} (2005), 13081-13086.
 
 \bibitem{Hell}
S. Hell and J. Wichmann, {\em Breaking the diffraction resolution limit by stimulated emission: stimulated-emission-depletion fluorescence microscopy},
 Opt. Lett., \textbf{19} (1994), 780-782.
 
 \bibitem{HGM}
S. Hess, T. Girirajan, and M. Mason, {\em Ultra-high resolution imaging by fluorescence photoactivation localization microscopy},  Biophys. J., \textbf{91} (2006), 4258-4272.

\bibitem{Huang}
F. M.  Huang, \textit{et al.}, {\em Nanohole array as a lens}, Nano Lett., \textbf{8} (2008), 2469-2472.

\bibitem{Lemoult-2011}
F. Lemoult, M. Fink, G. Lerosey,
{\em Acoustic resonators for far-field control of sound on a subwavelength scale},  Phys. Rev. Lett., \textbf{107} (2011), 064301.

\bibitem{LF}
W. Liao and A. Fannjiang, {\em MUSIC for single-snapshot spectral estimation: Stability and super-resolution}, Appl. Comput. Harmon. Anal., \textbf{40} (2016), 33-67.

\bibitem{lin_shipman_zhang}
J. Lin, S. Shipman, and H. Zhang,  {\em A mathematical theory for Fano resonance in a periodic array of narrow slits}, SIAM J. Appl. Math., to appear.

\bibitem{lin_zhang17}
J. Lin and H. Zhang, {\em Scattering and field enhancement of a perfect conducting narrow slit}, SIAM J. Appl. Math.,  \textbf{77} (2017), 951--976.

\bibitem{lin_zhang18_1}
J. Lin and H. Zhang, {\em Scattering by a periodic array of subwavelength slits I: field enhancement in the diffraction regime}, Multiscale Model. Simul., \textbf{16} (2018), 922--953.

\bibitem{lin_zhang18_2}
J. Lin and H. Zhang, {\em Scattering by a periodic array of subwavelength slits II: surface bound states, total transmission and field enhancement in the homogenization regimes},
 Multiscale Model. Simul., \textbf{16} (2018), 954--990.

\bibitem{lin_zhang19_1}
J. Lin and H. Zhang, {\em  An integral equation method for numerical computation of scattering resonances in a narrow metallic slit},
J. Comput. Phys., \textbf{385} (2019), 75-105.

\bibitem{lin_zhang19_2}
J. Lin and H. Zhang, {\em Mathematical analysis of surface plasmon resonance by a nano-gap in the plasmonic metal},
SIAM J. Math. Anal., \textbf{51} (2019), 4448-4489.

\bibitem{lin_oh_zhang20}
J. Lin, S-H. Oh, and H. Zhang, {\em Sensitivity of resonance frequency in the detection of thin layer using nano-slit structures},
submitted. 

\bibitem{lin_zhang20_1}
J. Lin and H. Zhang, {\em 
Fano resonance in metallic grating via strongly coupled subwavelength resonators},  European J. Appl. Math., to appear.

\bibitem{NB}
L. Novotny and B. Hecht, {\em Principles of Nano-Optics}, Cambridge University Press (2006).

\bibitem{Oh_Altug}
S. Oh and H. Altug, {\em Performance metrics and enabling technologies for nanoplasmonic biosensors}, Nat. Commun., \textbf{9} (2018), 5263.

\bibitem{Osher2}
S. Osher, \textit{et al.}, {\em An iterative regularization method for total variation-based image restoration}, Multicale Model. Simul., \textbf{4} (2005), 460-489.

\bibitem{Rayleigh}
L. Rayleigh, {\em On the theory of optical images with special reference to the optical microscope}, Phil. Mag., \textbf{5} (1896),
167-195.

\bibitem{RWCSF}
R. C. Reddick, \textit{et al.}, {\em Photon scanning tunneling microscopy}, Rev. Sci. Instrum., \textbf{61} (1990) 3669-3677.

\bibitem{ROF}
L. Rudin,  S. Osher, and E. Fatemi, {\em Nonlinear total variation based noise removal algorithms}, Physica D: Nonlinear Phenomena, \textbf{60} (1992): 259-268.

\bibitem{RBZ}
M. Rust, M. Bates, and X. Zhuang, {\em Stochastic optical reconstruction microscopy (STORM) provides sub-diffraction-limit image resolution}, Nat. Methods, \textbf{3} (2006), 793.

 \bibitem{XW}
Y. Xia and G. Whitesides, {\em Soft lithography}, Ann. Rev. Mater. Sci., \textbf{28} (1998), 153-184.

\bibitem{Liu-Zhang}
P. Liu and H. Zhang.
{\em Computational resolution limit: a theory towards super-resolution},
submitted.

\end{thebibliography}
\end{document}